\documentclass[journal]{IEEEtran}
\ifCLASSINFOpdf
\else
\fi
\usepackage[dvips,dvipdfm]{graphicx}
\usepackage{color}
\newcommand{\bea}{\begin{eqnarray}}
\newcommand{\eea}{\end{eqnarray}}
\newcommand{\eq}{&=&}
\newcommand{\nn}{\nonumber \\ }
\newcommand{\f}[2]{\frac{#1}{#2}}
\newcommand{\p}{\partial }
\newcommand{\pp}[2]{\frac{\p #1}{\p #2}}
\newcommand{\dis}[1]{\displaystyle{#1}}
\newcommand{\QC}{{Q_{c}}}
\newcommand{\QD}{{Q_{d}}}
\newcommand{\QV}{{Q_{v}}}
\newcommand{\F}{\overline{F}}
\newcommand{\Sref}[1]{eq. (\ref{#1})}
\newcommand{\Href}[1]{Fig. \ref{#1}}
\renewcommand{\a}{\alpha}
\renewcommand{\b}{\beta}
\renewcommand{\l}{\lambda}
\newcommand{\ve}{\varepsilon}
\renewcommand{\d}{\delta}
\newcommand{\s}{\sigma}
\begin{document}
\title{Improved and Developed Upper Bound of Price of Anarchy in Two Echelon Case}
\author{Takashi Shinzato 
        and Ikou Kaku
\\
Department of Management Science and Engineering\\ Graduate School of Systems Science and Technology, Akita Prefectual University\\
shinzato@akita-pu.ac.jp\qquad ikou$\_$kaku@akita-pu.ac.jp
}
\markboth{
}%
{Shell \MakeLowercase{\textit{et al.}}: Bare Demo of IEEEtran.cls for Journals}
\maketitle
\begin{abstract}
Price of anarchy, the performance ratio, which could characterize the loss of efficiency of the distributed supply chain management compared with the integrated supply chain management is discussed by utilizing newsvendor problem in single period which is well-known. In particular, some of remarkable distributed policies are handled, the performance ratios in each case which have been investigated in the previous works are analyzed theoretically and the tighter upper bound of price of anarchy and the lower bound are presented. Furthermore our approach is developed based on a generalized framework and a  geometric interpretation of price of anarchy is appeared via the literature of convex optimization.
\end{abstract}
\IEEEpeerreviewmaketitle
\section{Introduction}
\IEEEPARstart{M}{easuring} the efficiency of supply chains plays an important role in operations management because there exit many complicated factors (various contracts, order policies and physical structures) which could influence the decision making process. In practice, several formulae have been reported in most of the earlier studies \cite{Cachon2001,Chen,Tims,Perakis-Roels,Silver,Simchi}. In history, bullwhip effect, which could compare the variance of orders with that of demands, has been widely used to evaluate the efficiency of supply chain management in the stationary market 
 \cite{Chen}. Although it is straightforward to forecast the optimal order in the following term via finite instance of demand and order observed according to the manner of bullwhip effect, if the market behaves in equilibrium, it is too hard to determine the optimal order of goods in the typical case with respect to the given supply chain. While, as another approach, recently numerical experiments in Cachon (2004) indicate that the relative efficiency of a two-stage decentralized supply chain could be as low as 70-90$\%$ for either push or pull configurations with  price-only contract policy \cite{Cachon2004}. Perakis and Roels (2007)  mathematically extends Cachon's works into several different supply chain configurations such as push or pull inventory positioning, two or more stages, serial or assembly systems, single or multiple competing suppliers, and single or multiple retailers \cite{Perakis-Roels}. By introducing the concept of price of anarchy into supply chain, which can measure the ratio of the performance of the centralized system to the worst 
performance of the decentralized system (Koutsoupias and Papadimitriou 1999 and Papadimitriou 2001), they found that even in a two-stage supply chain the loss of efficiency might be more than 42$\%$ under the same conditions of price-only contract, and pointed that a pull inventory configuration generally outperforms a push configuration \cite{Koutsoupias1999,Papadimitriou2001,Simchi}. 

The contribution of Perakis and Roels (2007) is one of the most pioneer investigations in quantifying the performance of supply chain
. However, since their argument is hard to be guaranteed mathematically, rigorously and sufficiently anywhere, we need to improve several points in their logic for more practical use. In this paper, tighter upper bound of price of anarchy is theoretically proved and a lower bound of price of anarchy is presented firstly due to more accurate treatment. Furthermore a geometric interpretation of price of anarchy is also demonstrated which can give a clear illustration of the loss of efficiency of the given distributed supply chain. We only consider the both bounds of price of anarchy in two echelon case because of the most fundamental case, however it turns out that our approach would be simply applied in more complicated case.

This remainder of the present paper is organized as follows; in the next section, the newsvendor problem is presented for simplicity of our discussion, and the derived results in the previous work \cite{Perakis-Roels} is introduced briefly. Furthermore an improved upper bound and a lower bound of the loss of efficiency are explained. Section \ref{sec3} addresses a generalization of newsvendor problem and assesses the optimal inventory levels and the performance ratios in each mode. In Section \ref{sec4}, a geometric interpretation of price of anarchy is intuitively provided and we confirm that the performance ratio and the analytical procedure handled here are possible to be one of the most unbeatable frameworks. The final section is devoted to a summary and future work.
\section{Model setting\label{sec2}}
Concerned with distribution of goods and information sharing, supply chain management is one of the most vital interests in the cross-disciplinary fields. As one of the most pivotal topics, we discuss here the manner of the inventory management theoretically, in particular, (1) how to determine the optimal inventory level with respect to the given supply chain management and (2) how to assess the loss of efficiency of suboptimal policy. For simplicity of our argument, we restrict the model which is well-defined and is introduced below. Therefore applying  our demonstration, one could indeed improve and develop this approach so as to resolve more practical case. 
\subsection{Newsvendor problem}
Newsvendor problem is modeled as follows; if the firm prepares the inventory of goods $Q$ and the order in the market is $\xi$ , his profit in the single period is expected as follows;
\bea
\pi(\xi):\eq-cQ+p\min(Q,\xi)\label{eq1},
\eea
where the purchasing cost and the selling price describe $c$ and $p$, respectively, furthermore $\min(Q,\xi)$ denotes the lesser value of $Q$ and $\xi$. Note that $c\le p$ is needed in nature since the firm won't prefer to stock and buy the goods in the case $c>p$ (briefly the profit $\pi(\xi)\le0$ at $c>p$ in other words), and notice that $Q$ and $\xi$ are assumed as nonnegative and real numbers without the loss of generality. Here the opportunity loss is not handled and the inventory space is large enough as a matter of convenience, however it turns out that our approach would be simply applied in the case with opportunity loss and the upper bound of inventory level. Generally speaking, it is too hard to estimate that the demand in each single period is fixed. Hence, let us propose that the demand $\xi$ is stochastically drawn from the given density function $f(\xi)$ with the cumulative probability. Now, the expected aggregate profit is represented as follows;
\bea
\Pi:=
\int_0^\infty d\xi f(\xi)\pi(\xi)
=-cQ+p\int_0^Qd\xi\overline{F}\left(\xi\right)\label{eq2},
\eea
where $\F(\xi):=\int_\xi^\infty dxf(x)$ describes the cumulative probability whose stochastic variable is greater than or equal to $\xi$ and $\Pi$ is a concave function of the inventory quantity strictly (show appendix \ref{appag}). Note that $\Pi=0$ at $Q=0$ is required for any distribution of demand by definition and it implies that no stock is no benefit. Moreover as trivial, the supremum of the expected entire benefit $\Pi$ is indeed greater than or equal to zero because of the previous notice.

The production planner's purpose in general is to maximize his expected gross benefit $\Pi$ by adjusting the inventory level $Q$ in typical situation. However, nowadays the logistics, the distribution of goods in supply chain behaves like bloodstream in the human society, the necessity not only of the integrated inventory management but also of the distributed inventory management has been recognized. In the given distributed configurations, the problem who makes to store the inventory and/or who needs to decide the wholesale price in supply chain is one of the most vital issues, furthermore it is also important how one examines the loss of efficiency of the distributed management in some of remarkable configurations compared to the integrated management. 
\subsection{Centralized supply chain}
Let us review here the problem how the inventory level $Q$ is derived in order to maximize the expected entire profit of the given integrated (or centralized) supply chain. From \Sref{eq2}, the unique optimal solution which can be desired is intuitively derived as follows; $
\QC:=
\F^{-1}\left(r\right)
$ 
 where $r:=c/p$ and $\F^{-1}(y)(=x)$ represents the inverse function of $\F(x)(=y)$. Since the second term in \Sref{eq2} is a concave function of $Q$ (show appendix \ref{appd}), it turns out that the unique optimal solution is strictly determined.
\subsection{The profits in two echelon case}
The counterpart of the centralized case in supply chain management, the distributed (or decentralized) supply chain management is explained here. In the distributed case the expected profit could be divided into two distinguished profit functions as follows;
$\Pi=\Pi^{\rm M}+\Pi^{\rm R}$ in push serial supply chain 
and $\Pi=\Xi^{\rm M}+\Xi^{\rm R}$ in pull serial supply chain. Firstly, $\Pi^{\rm M}:=(w-c)Q$ and $\Pi^{\rm R}:=-wQ+p\int_0^Qd\xi\overline{F}\left(\xi\right)$ indicate the manufacturer's entire profit and the retailer's whole benefit, respectively, in the case that the retailer makes to stock the inventory 
in push serial supply chain. While $\Xi^{\rm M}:=-cQ+w\int_0^Qd\xi\overline{F}\left(\xi\right)$ and $\Xi^{\rm R}:=(p-w)\int_0^Qd\xi\overline{F}\left(\xi\right)$ describe the manufacturer's aggregate profit and the retailer's total benefit, respectively, in the situation that the manufacturer makes to store the inventory 
in pull serial supply chain. 
In each case, the leader decides the wholesale price $w$ (note that $c\le w\le p$ is required in practice, because these profits are satisfied with positivity) in order to maximize the leader's expected whole profit, while the follower should choose selfishly the optimal inventory level $Q$ by employing the optimization problem of the follower's expected benefit with respect to the given wholesale price, that is, it is the scenario  of Stackelberg leadership game \cite{Roughgarden}
. In the previous investigations, the analysis of the class of increasing generalized failure rate distribution, where one ensures that the optimization problem in the decentralized configurations could possess the well-defined optimal solution, has been reported comparatively well with regard to several concrete distributed configurations in two echelon case as follows;
\begin{description}
\item[{ (a) }]The manufacturer is the decision maker in push serial supply chain. $\QD$ is determined by the following equation; $
\F\left(\QD\right)\left(1-g\left(\QD\right)\right)=r$ where 
$g(Q):=\f{Qf(Q)}{\F(Q)}\label{eq13}$
is utilized. In this paper $g(Q)$ is assumed as a nondecreasing function of $Q$ and $0\le g(Q)\le1$ because it is guaranteed that the optimal solution of the follower's optimization problem is unique. Thus $g(Q)$ is termed as increasing generalized failure rate. In addition, the desirable wholesale price is yielded as $w=p\F(\QD)$.
\item[{ (b) }]The retailer is the decision maker in push serial supply chain. $\QD$ is consistent with the inventory level in the integrated supply chain management, because the inventory is stored at the downstream site and it is to be expected  that the wholesale price is equal to the purchasing cost. 
\item[{ (c) }]The manufacturer is the decision maker in pull serial supply chain. $\QD$ is also consistent with the inventory level in integrated supply chain, because the inventory is stored at the leader's site and it is to be desired that the wholesale price corresponds to the selling price. 
\item[{ (d) }]The retailer is the decision maker in pull serial supply chain. $\QD$ is decided by the following equation; 
$\F\left(\QD\right)\left(1+l\left(\QD\right)\right)^{-1}=r$ where 
$l(Q):
=\f{\dis{f(Q)}}{\dis{\F^2(Q)}}\int_0^Qd\xi\F(\xi)
$ 
is employed and the optimal wholesale price is derived as $w=c/\F(\QD)$. If $g(Q)$ is increasing generalized failure rate, then $l(Q)$ is nondecreasing function of $Q$ strictly \cite{Cachon2004,Perakis-Roels}. 
\end{description}
In practice, the production planner should choose the most appropriate distributed inventory management in some remarkable configurations by some means. For that reason, our purpose is here to examine the following measurement so as to assess the loss of efficiency of each configuration. Price of anarchy, the performance ratio which 
can characterize the loss of efficiency of the expected whole benefit of the given decentralized inventory management compared with that of the given centralized inventory management and which is well-known, is denoted as follows;
\bea
{\rm PoA}:\eq\f{\dis{-c\QC+p
\int_0^{\QC}d\xi\overline{F}\left(\xi\right)}}
{\dis{-c\QD+p
\int_0^{\QD}d\xi\overline{F}\left(\xi\right)}},
\eea
where \Sref{eq2} is utilized and it turns out that ${\rm PoA}$ is greater than or equal to unit in general (show appendix \ref{appa}). 
\subsection{Rigorous results derived in the previous works}
With respect to the ensemble of increasing generalized failure rate, the rigorous results of price of anarchy were presented in the previous work as follows \cite{Perakis-Roels}; (a) ${\rm PoA}\le(1-k)^{-\f{1}{k}}-(1-k)^{-1}$, (b) ${\rm PoA}=1$, (c) ${\rm PoA}=1$ and (d) ${\rm PoA}\le(1+l)^{1+\f{1}{l}}-(1+l)$, 
where $k:=g\left(\QD\right)$ and $l:=l\left(\QD\right)$ are utilized (the indices are mentioned in the previous subsection). Indeed, although the previous work explained that the upper bounds in (a) 
and (d) 
are comparatively tight, it is hard to assess the loss of efficiency in precision and in good faith, therefore we discuss more conscientiously and obtain a tighter upper bound compared with the derived one in Perakis and Roels (2007), and a tighter lower bound with respect to the class of increasing generalized failure rate exactly.
\subsection{An upper and lower bound of cumulative probability}
From the definition of increasing generalized failure rate, 
$\log\F(\xi)=
-\int_0^\xi dy\f{g(y)}{y}$
is analytically yielded where $\F(0)=1$. Moreover with regard to the ensemble of increasing generalized failure rate,
$\max\left(\F\left(\QC\right),\F\left(\QD\right)\left(\QD\right)^s\xi^{-s}\right)\le\F\left(\xi\right)
\le\max\left(\F\left(\QC\right),\F\left(\QD\right)\left(\QD\right)^k\xi^{-k}\right)
$ 
 is obtained where $\QD\le \xi\le\QC$ and $k=g\left(\QD\right)$ and $s:=g\left(\QC\right)$ are employed.  
According to the ratio of the inventory level in the integrated case to the one in the distributed case, $\a:=\QC/\QD$, we obtain an upper bound and a lower bound of the integration as follows;
\bea
{\cal L}(\a,s)\le\dis{\int_{\QD}^{\QC}d\xi\F\left(\xi\right)}\qquad\qquad
\qquad\qquad\qquad\qquad\ 
\nn
\le\left\{\begin{array}{ll}
\QD\F\left(\QD\right)\f{
\a^{1-k}-1
}{1-k}&(1-k)^{-\f{1}{k}}\le\a\\
{\cal L}(\a,k)&
(1-k)^{-\f{1}{s}}\le\a\le(1-k)^{-\f{1}{k}}
\end{array}\right.
\label{eq23}
\eea
where ${\cal L}(\a,t):=\QD\F\left(\QD\right)\left[(1-k)\a+\f{t\left(1-k\right)^{1-\f{1}{t}}-1}{1-t}\right]$~is~used. Notice that $\a<(1-k)^{-\f{1}{s}}$ is not satisfied with the boundary condition, $\F(\QC)=r$ and ${\cal L}(\a,t)$ is a nonincreasing and convex function of $t$ for any $\a$. Since the integration in \Sref{eq23} is evaluated more accurately compared with their insufficient discussion in \cite{Perakis-Roels}, the comparative tight upper bound of price of anarchy and the lower bound are expected fortunately. Furthermore, since the cumulative probability is a nonincreasing function, the upper bound in this interval $\QD\le\xi$ is not possible to exceed the probability at $Q=0$ in nature, the inequality 
$\F\left(\QD\right)\le\F(\xi)\le\min\left(1,\F\left(\QD\right)\left(\QD\right)^k\xi^{-k}\right)$ 
is derived, and an upper bound and a lower bound of the integration, 
\bea
&&
\QD\F\left(\QD\right)\le\int_0^{\QD}d\xi\F(\xi)\nn
&\le& \QD\F\left(\QD\right)\left[1+\f{k\left(1-\F^{\f{1}{k}-1}\left(\QD\right)\right)}{1-k}\right] ,\label{eq27}
\eea
is obtained. 
\subsection{Both bounds of price of anarchy; the manufacturer is the leader in push serial supply chain}
\begin{figure}[hbt]
\begin{center}
\includegraphics[width=8cm,height=6cm]{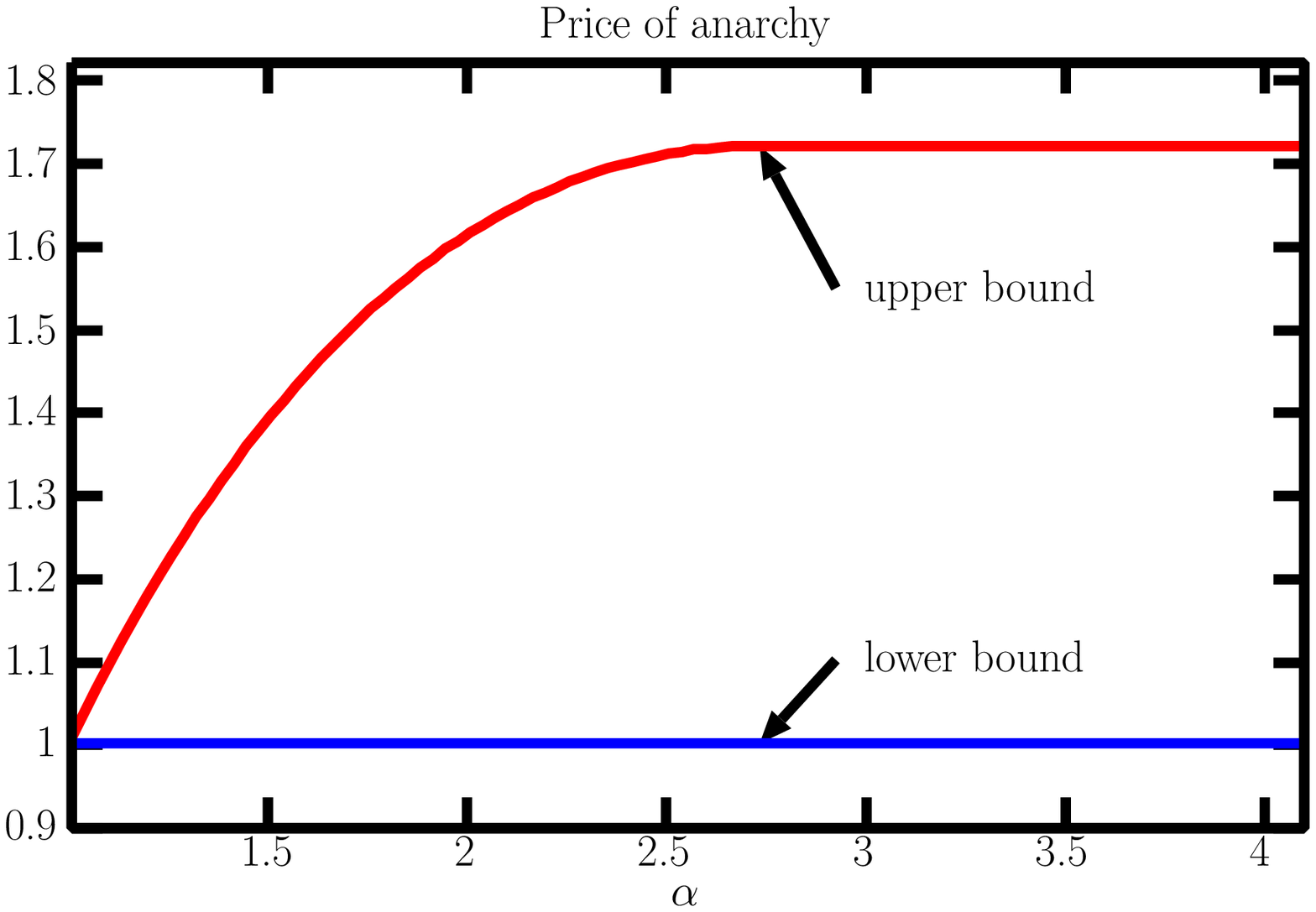} 
\caption{\label{poa001}The ratio $\a$ v.s. both bounds of price of anarchy at $k=0.01$, $\F(\QD)=0.5$ and $s\simeq1.0$} 
\includegraphics[width=8cm,height=6cm]{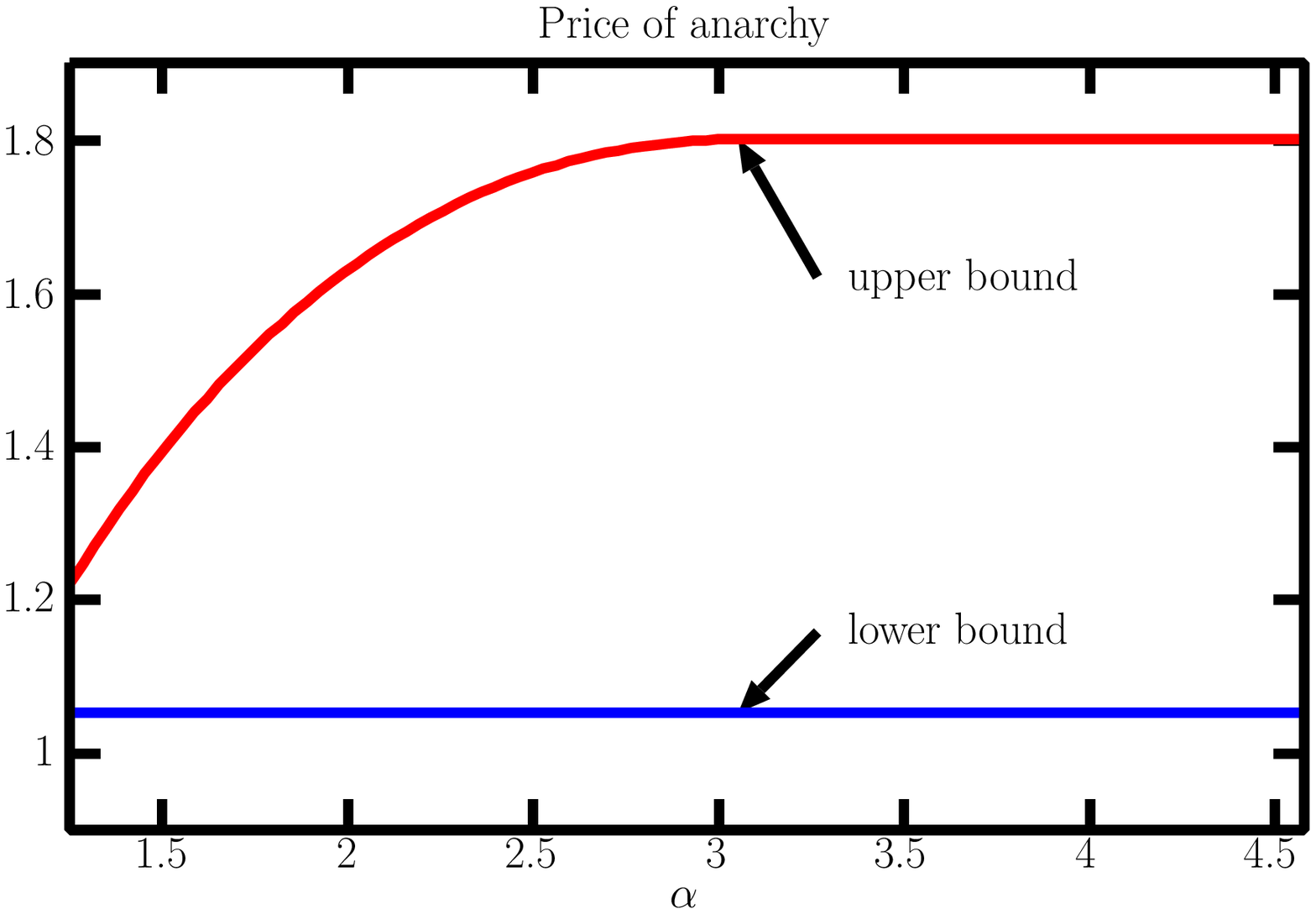} 
\caption{\label{poa020}The ratio $\a$ v.s. both bounds of price of anarchy at $k=0.20$, $\F(\QD)=0.5$ and $s\simeq1.0$} 
\end{center}
\end{figure}

According to the argument in the previous work \cite{Perakis-Roels}, 
by definition, one can replace price of anarchy as follows;
${\rm PoA}=
1+\f{{\int_{\QD}^{\QC}d\xi\left(\F(\xi)-r\right)}}{{-r\QD+\int_0^{\QD}d\xi\F(\xi)}}$.
Thus employing \Sref{eq23} and \Sref{eq27}, an upper bound
\bea
{\rm PoA}\le\left\{\begin{array}{ll}
(1-k)^{-\f{1}{k}}-(1-k)^{-1}&(1-k)^{-\f{1}{k}}\le\a\\
\f{\a^{1-k}-(1-k)^2\a}{k(1-k)}-(1-k)^{-1}&{\rm otherwise}
\end{array}\right.
\label{eq171}
\eea
and a lower bound
\bea
{\rm PoA}\ge
\f{\dis{\f{s(1-k)^{1-\f{1}{s}}-s}{1-s}-\f{\dis{kr^{\f{1}{k}-1}(1-k)^{1-\f{1}{k}}-k}}{\dis{1-k}}}}{k+\f{\dis{k-kr^{\f{1}{k}-1}(1-k)^{1-\f{1}{k}}}}{\dis{1-k}}},
\eea
are simply yielded. It turns out that the numerator of the first term of the upper bound in $(1-k)^{-\f{1}{s}}\le\a\le(1-k)^{-\f{1}{k}}$ in \Sref{eq171} is a nondecreasing and concave function of $\a$. The supremum of the right hand side in \Sref{eq171} was already presented in \cite{Perakis-Roels}, while the comparative tight upper bound is to be desired in $(1-k)^{-\f{1}{s}}\le\a\le(1-k)^{-\f{1}{k}}$. Fig. \ref{poa001} shows that in the limit of $k\to0$ for any $s>k$, both bounds at $\a\to(1-k)^{-\f{1}{s}}$ are close to unit and if $s=k$, the lower bound $\left(1+\f{2-k}{\left(1-r^{\f{1}{k}-1}\right)(1-k)^{1-\f{1}{k}}}\right)^{-1}$ is greater than or equal to unit at $\a\ge(1-k)^{-\f{1}{k}}$. While \Href{poa020} indicates the behavior at $k=0.20$ and $r=0.40$ and it is found that both bounds are greater than unit at any ratio $\a$.
\subsection{Both bounds of price of anarchy; the retailer is the leader in pull serial supply chain}
In this case, 
 compared with the previous subsection, $k$ and $s$ are rewritten as $1-k=(1+l)^{-1}$ and $1-s=(1+t)^{-1}$, respectively, where $l=l\left(\QD\right)$ and $t\ge l$, then an upper bound and a lower bound;
\bea
{\rm PoA}&\le&
\left\{\begin{array}{ll}
(1+l)^{1+\f{1}{l}}-(1+l)&(1+l)^{1+\f{1}{l}}\le\a\\
\dis{\f{(1+l)^2\a^{\f{1}{1+l}}-\a}{l}-(1+l)}
&{\rm otherwise}
\end{array}\right.\nn
{\rm PoA}&\ge&\f{t(1+l)^{\f{1}{t}}-t+l-lr^{\f{1}{l}}(1+l)^{\f{1}{l}}}{(1+l)^{-1}+
l-lr^{\f{1}{l}}(1+l)^{\f{1}{l}}}\nonumber
\eea
are also yielded.
\subsection{Example: Nonnegative order drawn from normal Gaussian distribution}
\begin{figure}[h]
\begin{center}
\includegraphics[width=8cm,height=6cm]{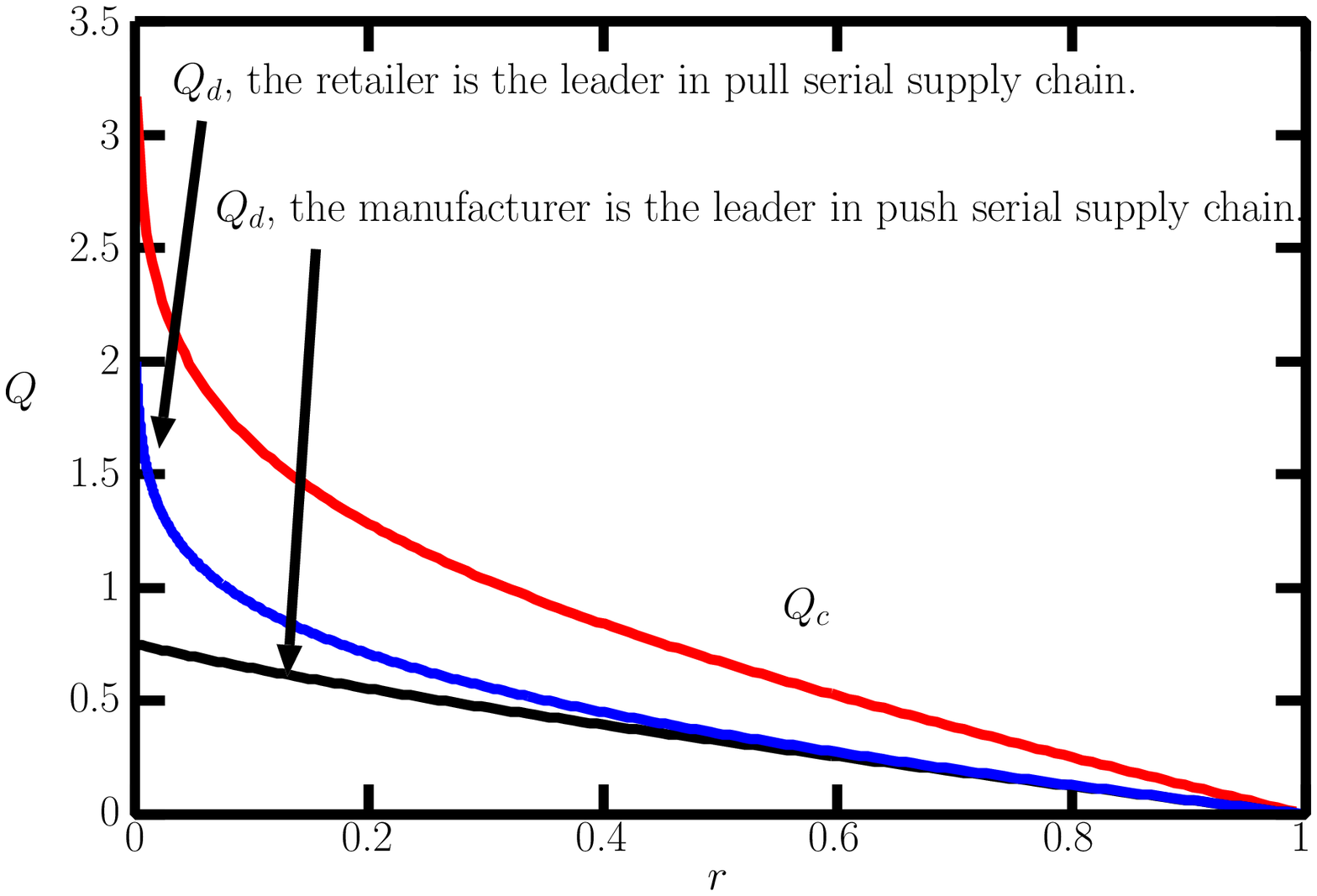}
\caption{\label{1fig}The ratio $r=c/p$ v.s. $\QC$ and $\QD$.}
\includegraphics[width=8cm,height=6cm]{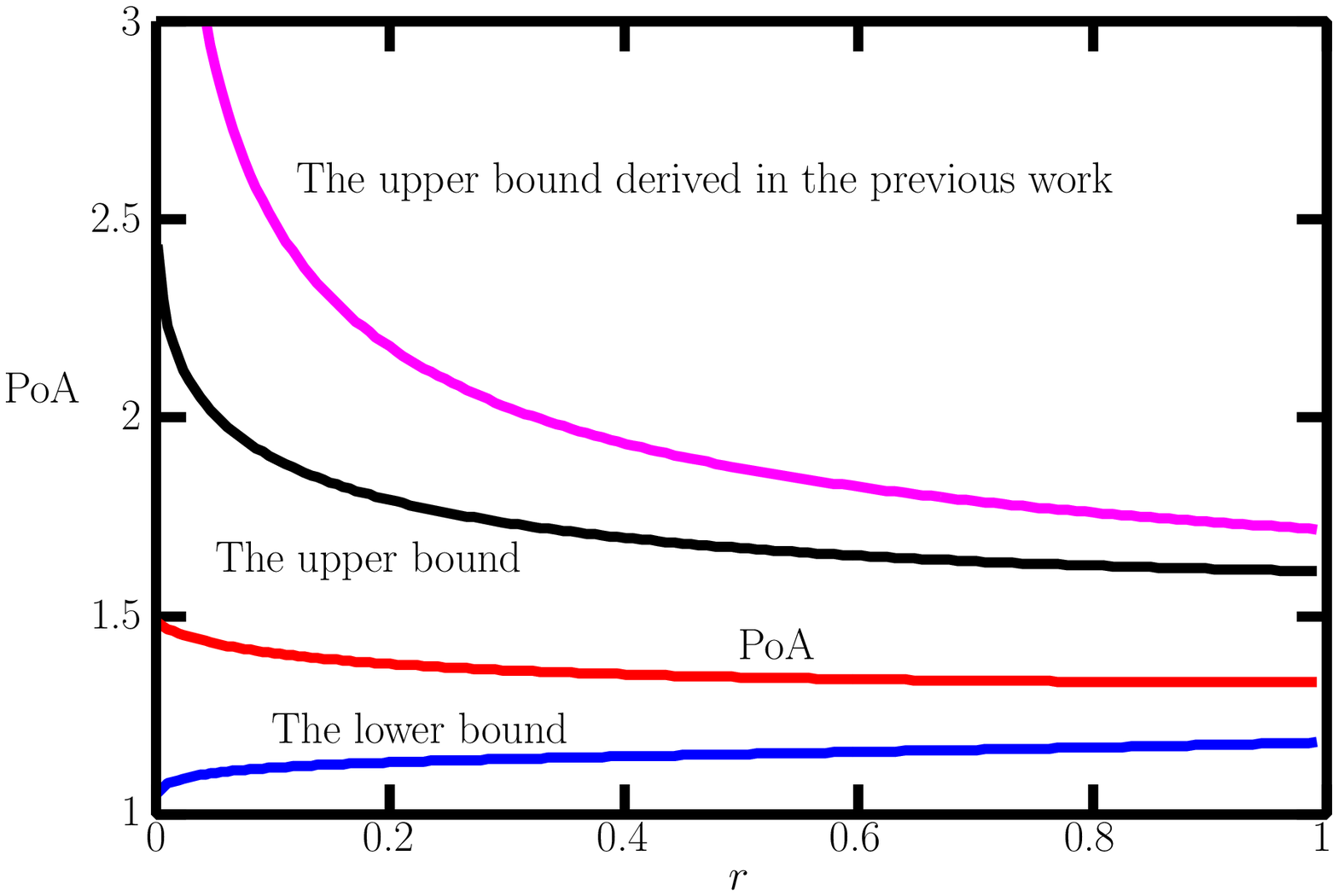}
\caption{\label{2fig}The ratio $r$ v.s. price of anarchy and the bounds. In push serial supply chain, the manufacturer is the decision maker.}
\includegraphics[width=8cm,height=6cm]{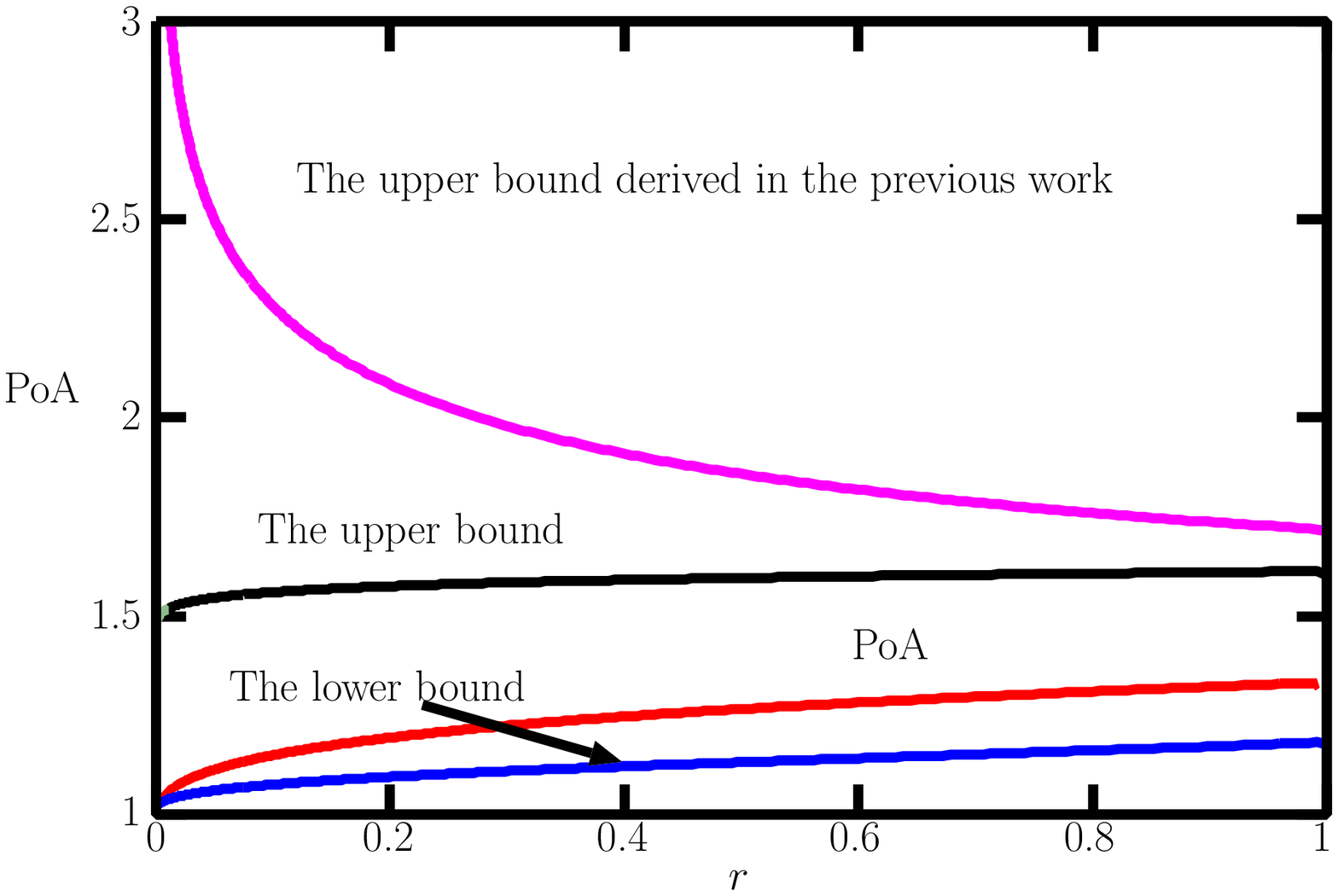}
\caption{\label{3fig}The ratio $r$ v.s. price of anarchy and the bounds. In pull serial supply chain, the retailer is the decision maker.}
\end{center}
\end{figure}

Let us confirm the effectiveness of our approach with a novel toy model. 
For simplicity, it is assumed that the demands are independently and identically distributed according to most of the previous works. Here the density function of demand, $f(\xi)$ is satisfied with $\f{2}{\sqrt{2\pi}}e^{-\f{\xi^2}{2}}$ for $\xi\ge0$ and $0$ otherwise. One can easily validate that $g(Q)$ of this model is increasing generalized failure rate. Thus $\QC$ and $\QD$ in the two cases of ${\rm PoA}\ne1$, are illustrated in \Href{1fig}. Furthermore, as shown in \Href{2fig} and in \Href{3fig} that the numerical results of ${\rm PoA}$ and the derived bounds are compared with each other in the case that the manufacturer is the decision maker in push serial supply chain and in the case that the retailer is the decision maker in pull serial supply chain, respectively. In conclusion, it turns out that our approach is valid in this model.
\subsection{Example: Autoregression model with $\chi$ square noise}
\begin{figure}[hbt]
\begin{center}
\includegraphics[width=8cm,height=6cm]{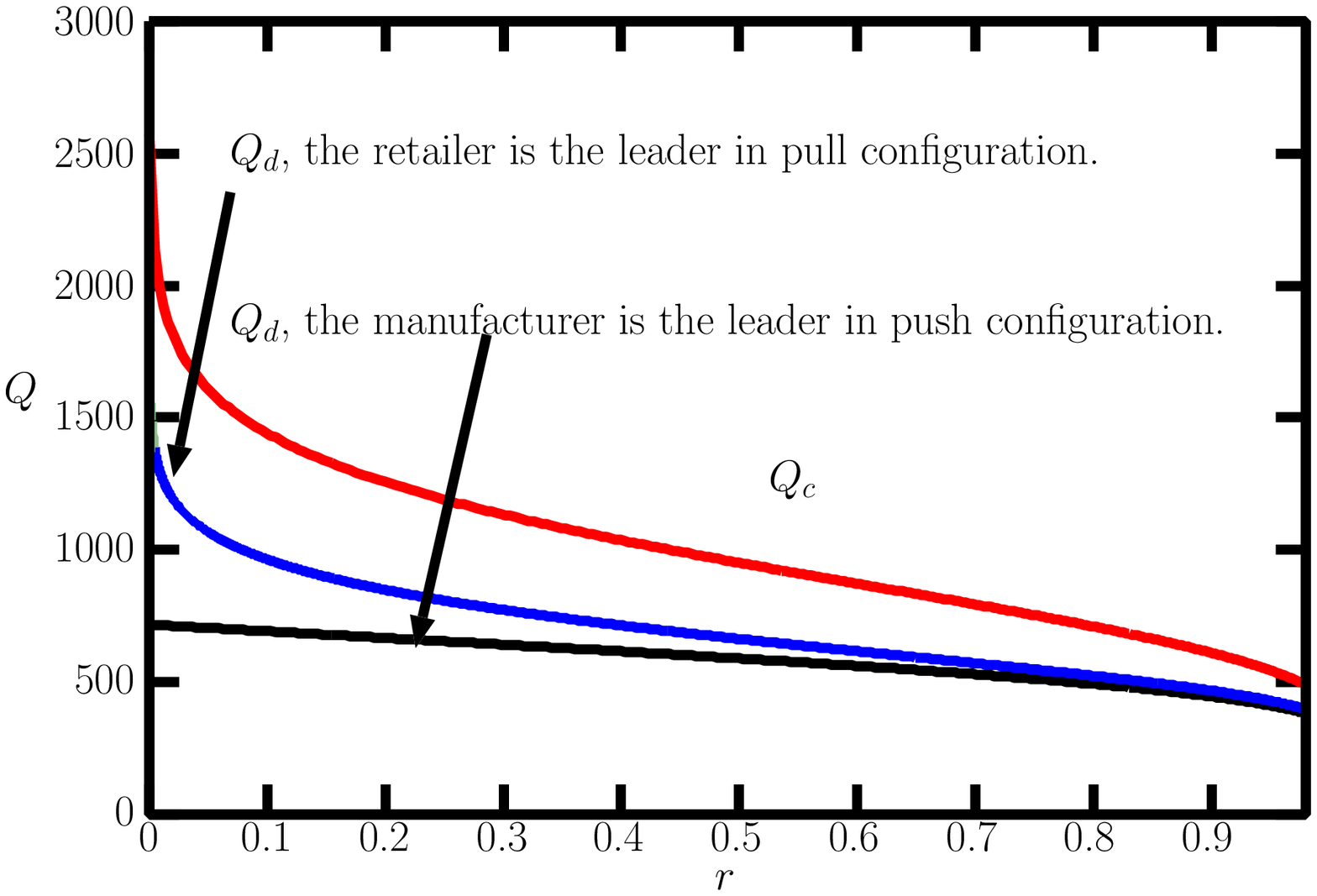} 
\caption{\label{chi-time-series2}The ratio v.s. the inventory levels at $\b=0.9$ and $\s^2=100.0$.}
\includegraphics[width=8cm,height=6cm]{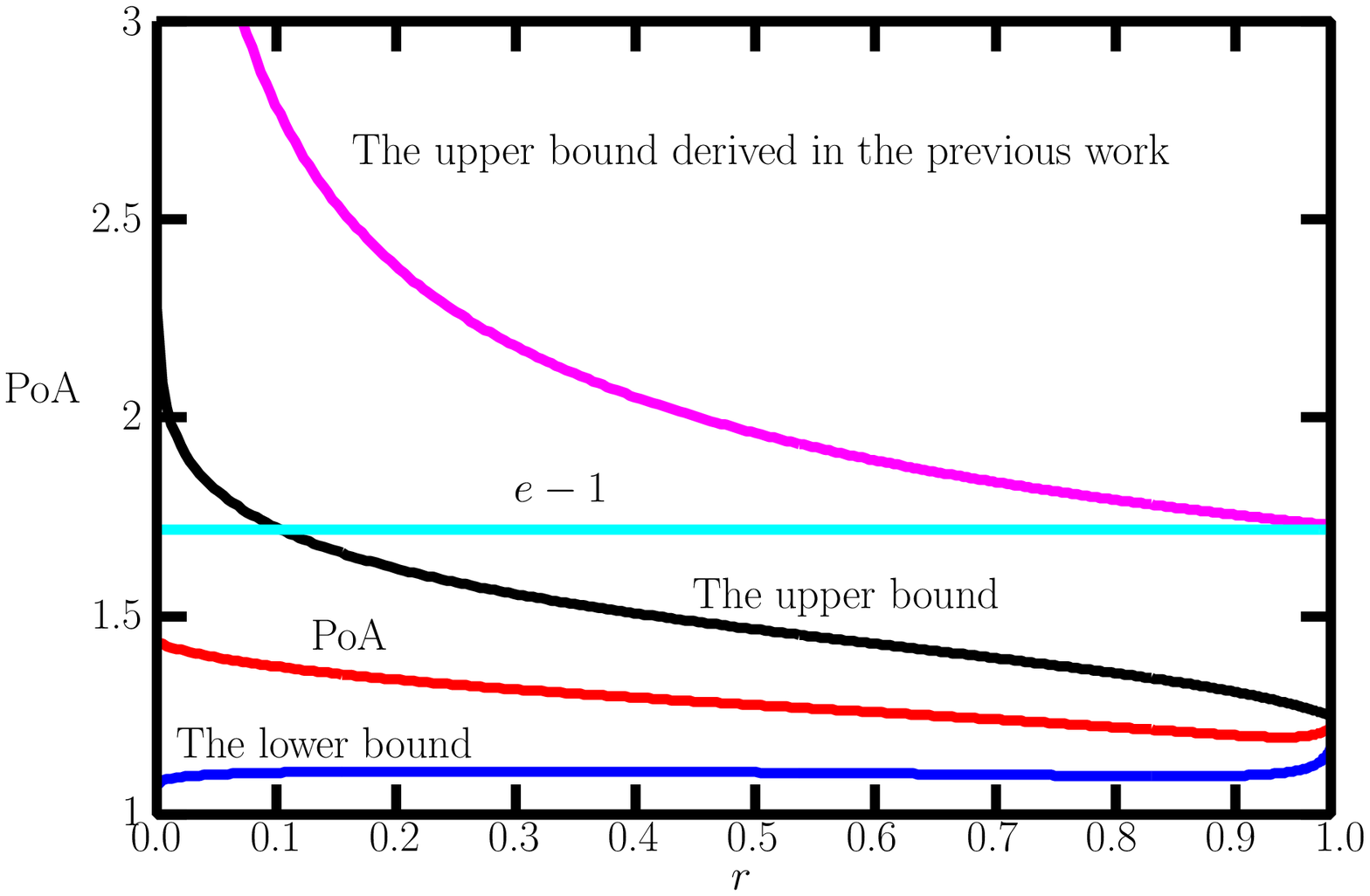} 
\caption{\label{chi-time-series3}The ratio $r$ v.s. price of anarchy and the derived bounds. In push serial supply chain, the manufacturer is the decision maker.}
\includegraphics[width=8cm,height=6cm]{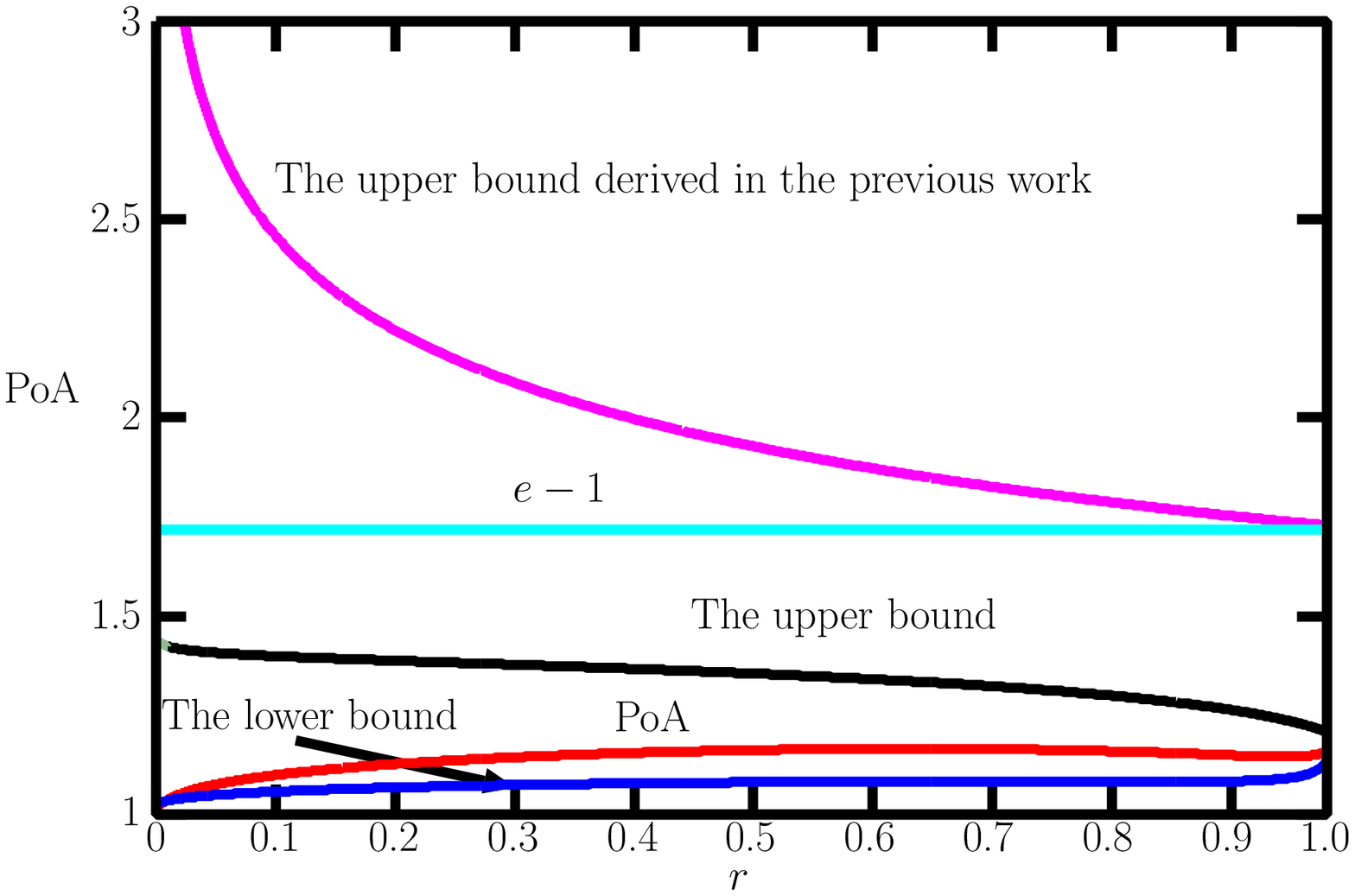} 
\caption{\label{chi-time-series4}The ratio $r$ v.s. price of anarchy and the derived bounds. In pull serial supply chain, the retailer is the decision maker.}
\end{center}
\end{figure}

As the real-world data analysis, it is natural that the demands are correlated with one another, for instance, a market trend, rather than that they are independently and identically distributed  \cite{Kabashima2008,Shinzato2008,Shinzato2009}. However, in most of the earlier studies, for simplicity, the demands were independently and identically distributed and this assumption is obviously not practical since real-world date are usually somewhat biased and correlated across the instance. In general, the demand might consist of the trend effect which implies correlation and an uncertain factor which connotes noise (c.f. bullwhip effect). Therefore as the first step, let us handle a novel toy model in order to confirm whether or not these bounds of price of anarchy which have been presented in this paper are influenced by the correlation in the market. Well, the demand at discrete time $T$ (denoted by $\xi_T$) is generated from the following autoregression model; 
$\xi_{T+1}=\b \xi_T+\s^2\chi_T^2$, 
where $\b$ indicates a dumping coefficient in $0\le\b\le1$ and $\chi_T^2$, the random variable at time $T$, is drawn from $\chi$ square distribution with one degree of freedom (in addition, for simplicity each noise is assumed to be independently and identically distributed and $\s^2$ implies a noise intensity).  Although there is not an aggressive premise, because the demand is guaranteed to be always nonnegative, this toy model is accepted here. 
\begin{figure}[hbt]
\begin{center}
\includegraphics[width=8cm,height=6cm]{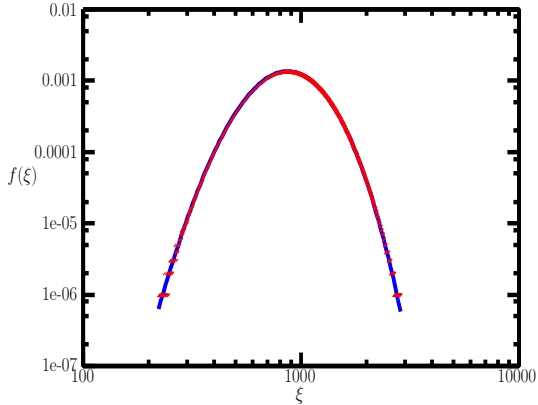} 
\caption{\label{poa00}It is well-known that the density function $f(\xi)$ is analytically yielded via the cumulant generating function, the logarithm function of characteristic function, $\log\int_0^\infty d\xi f(\xi)e^{i\theta\xi}=-\f{1}{2}\sum_{k=1}^\infty\log\left(1-2i\theta\s^2\b^k\right)$ \cite{Feller}. However we estimate here the density function utilizing the histogram numerically evaluated by $0.5\times 10^9$ demands which are consisted of the sequences which are randomly chosen from the times series, which is stable, and we fit the logarithm of the density function derived numerically into the fifth-degree polynomial function of $\log\xi$, which is supported by minimization of the the leave-one-out cross validation error \cite{Shinzato2009}. $f(\xi)$ at $\b=0.9$ and $\s^2=100.0$ is illustrated in this figure.
}
\end{center}
\end{figure}
Now, given $\b$ and $\s^2$, and the density function of demand $f(\xi)$ is stable, one can evaluate the loss of efficiency, ${\rm PoA}$ with respect to the ratio $r=c/p$, furthermore, we can confirm whether or not both bounds are valid. The density function of demand, $f(\xi)$ at $\b=0.9$ and $\s^2=100.0$ is illustrated in \Href{poa00}, $\QC$ and $\QD$ in the two cases are depicted in \Href{chi-time-series2}. As shown in \Href{chi-time-series3} and in \Href{chi-time-series4} that the numerical result of ${\rm PoA}$ and the upper and lower bounds are compared with one another at $\b=0.9$ and $\s^2=100.0$ in the case that the manufacturer is the leader in push serial supply chain and in the case that the retailer is the leader in pull serial supply chain, respectively. Likewise, it turns out that our procedure is validly supported in this correlated model.

\section{Generalized Framework\label{sec3}}
We have discussed price of anarchy, the performance ratio which can characterize the loss of efficiency of the given distributed supply chain management compared with the integrated management policy, based on newsvendor problem which is well-known. In particular, some of remarkable distributed policies are handled, the performance ratios in each case, for instance, who makes to store the inventory and/or who needs to decide the wholesale price, are analyzed theoretically, and the tighter upper bound of price of anarchy and the lower bound are presented. However, two points should be noted here that these results are not restricted to newsvendor problem (hereafter it is termed as original newsvendor problem) and price of anarchy is possible to be regarded as one of the most unbeatable feature quantity with respect to a broad class of optimization problems.

In this section, let us propose a generalization of the approach treated in the previous section (hereafter generalized newsvendor problem is handled intuitively, however it is also demonstrated in the literature of supply chain management as a matter of convenience). According to our argument, one could replace the profit function in \Sref{eq1} as $\pi(\xi)=-cQ+pm(\xi,Q)$, i.e. $m(\xi,Q)=\min(\xi,Q)$; it turns out that $m(\xi,Q)$ can describe a more complicated policy of price contracts, which could sufficiently characterize a market trend, for instance, the order of hard-to-find items in practice. Now, for simplicity of the discussion, we prepare the expected aggregate profit,
\bea
\label{eq37}\Pi:\eq-cQ+p{\cal M}(Q),
%
\eea
where ${\cal M}(Q):=\int_0^\infty d\xi f(\xi)m(\xi,Q)$ and ${\cal X}(Q):=\pp{{\cal M}(Q)}{Q}$ are defined. In order to extend the method, we should explain $m(\xi,Q)$, ${\cal M}(Q)$ and ${\cal X}(Q)$ simply. Firstly $m(\xi,Q)$ is here an arbitrary function which is satisfied with the following conditions, $\pp{^2{\cal M}(Q)}{Q^2}=\pp{{\cal X}(Q)}{Q}\le0$ and ${\cal M}(0)=0$. 
Because 
it  is required for the concavity of the expected total profit function (one would not presume that $m(\xi,Q)$ is a concave function of $Q$ in general, i.e. $m(\xi,Q)=\min(\xi,Q)$) and ${\cal M}(0)=0$ is necessary so as to connote that no stock is no benefit. Furthermore from ${\cal X}(Q)$ and ${\cal M}(0)=0$, ${\cal M}(Q)$ 
 is possible to be replaced as 
${\cal M}(Q)=
\int_0^Qd\xi{\cal X}(\xi)$ (it is not required to disclose the density function of demand in generalized newsvendor problem), 
where ${\cal X}(\xi)$ can be regarded as $\F(\xi)$ in the previous section. Further it implies that ${\cal X}(Q)$ is assumed as positive because the optimal inventory level is nonnegative. Lastly, $c\le p{\cal X}(0)$ is needed in nature (briefly $\Pi\le0$ for any inventory level is satisfied at $c\ge p{\cal X}(0)$).
\subsection{Centralized case and profit functions}
The unique optimal solution in the integrated case of generalized newsvendor problem is denoted as follows; $\QC={\cal X}^{-1}(r)$, 
where $r=c/p$ (notice that $0\le\f{c}{p{\cal X}(0)}\le1$) and ${\cal X}^{-1}(y)(=x)$ represents the inverse function of ${\cal X}(x)(=y)$. 
Next, the expected profit function $\Pi$ in \Sref{eq37} is  possible to be divided into two distinguished parts in push and pull configurations as follows; (i) $\Pi^{\rm M}:=(w-c)Q$, (ii) $\Pi^{\rm R}:=-wQ+p{\cal M}(Q)$, (iii) $\Xi^{\rm M}:=-cQ+w{\cal M}(Q)$ and (iv) $\Xi^{\rm R}:=(p-w){\cal M}(Q)$, 
where the wholesale price $w$ is satisfied with $c\le w\le p{\cal X}(0)$ in push serial supply chain and $c/{\cal X}(0)\le w\le p$ in pull serial supply chain, respectively. According to the previous argument, price of anarchy is also denoted as follows;
\bea
{\rm PoA}:=\f{\dis{-c\QC+p{\cal M}(\QC)}}
{\dis{-c\QD+p{\cal M}(\QD)}}.
\eea
In the follows, we explain how the optimal inventory level and the desirable wholesale price of each decentralized case are determined so as to measure the loss of efficiency.
\subsection{Manufacturer is the leader in push serial supply chain}
It is well-known that ${\rm PoA}$ in this situation of original newsvendor problem is not always equal to unit. In practice one can derive the optimal solution iteratively via the saddle point equation as follows;
\bea
\label{eq48}
\QD={\cal X}^{-1}(\ve)\ \ \left.\begin{array}{cc}
\Longrightarrow\\
\Longleftarrow
\end{array}
\right.\ \ 
\ve= r+g(\QD){\cal X}(\QD)
\eea
where $\ve:=w/p$ and a novel function, 
$g(Q):=-Q\pp{}{Q}\log{\cal X}(Q)$ 
are already used. By definition, $g(Q)$ is a nonnegative function of $Q$ in general. Furthermore in order to determine the unique solution of the decentralized management, we assume that $g(Q)$ is increasing generalized failure rate (note that if $g(Q)$ is termed as increasing generalized failure rate, then $g(Q)$ is satisfied with $0\le g(Q)\le1$ and $\pp{g(Q)}{Q}\ge0$).

In addition, although this iteration connotes the recursive procedure in order to resolve the optimal inventory level systematically in the distributed system, using \Sref{eq48}
, one can represent also the following relation as 
${\cal X}(\QD)\left(1-g(\QD)\right)=r$, 
where the desirable wholesale price is $w=p{\cal X}(\QD)$ by the definition of $\ve$.
\subsection{Retailer is the leader in push serial supply chain}
In this case, ${\rm PoA}$ is equal to unit because the derivative of the leader's expected profit $\Pi^{\rm R}$ with respect to the wholesale price $w$ is nonpositive, the optimal wholesale price is desirable to be consistent with the purchasing cost in $c\le w\le p{\cal X}(0)$, therefore it is possible to be regarded as the integrated system as the follower's benefit is zero, that is, $\QD$ is equivalent to $\QC$.

\subsection{Manufacturer is the leader in pull serial supply chain}
Fortunately, ${\rm PoA}$ coincides with the value derived in original newsvendor problem because the derivative of the leader's expected profit $\Xi^{\rm M}$ with respect to the wholesale price $w$ is nonnegative, the optimal wholesale price is desirable to be equal to the selling price in $c/{\cal X}(0)\le w\le p$, therefore it is possible to be regarded as the integrated system as the follower's benefit is zero, that is, $\QD$ is consistent with $\QC$.

\subsection{Retailer is the leader in pull serial supply chain}
It is sure that ${\rm PoA}$ in the last case of original newsvendor problem is not less than unit.
With respect to generalized newsvendor problem, let us evaluate the optimal solution sequentially utilizing the steepest descent method as follows;
\bea
\label{eq52}\QD={\cal X}^{-1}(\d)\ \ \left.\begin{array}{cc}
\Longrightarrow\\
\Longleftarrow
\end{array}
\right.
\ \ 
\f{1}{\d}=\f{1}{r}-\f{l(\QD)}{{\cal X}(\QD)},
\eea
where $\d:=c/w$ and a novel function $l(Q):
=-\f{{\pp{}{Q}\log{\cal X}(Q)}}
{{\pp{}{Q}\log{\cal M}(Q)}}$ 
are employed. Under the definition, $l(Q)$ is a nonnegative function of the inventory level in nature. Moreover, so as to derive the unique solution of the distributed management, we require that $l(Q)$ is a nondecreasing function of $Q$ as the sufficient condition (if $g(Q)$ is assumed as increasing generalized failure rate, then $l(Q)$ is a nondecreasing function of $Q$, show appendix \ref{appc}).

Additionally, although this iteration derived here implies the algorithmic procedure so as to assess the optimal inventory level in the decentralized system, applying \Sref{eq52}
, the following relation,
${\cal X}(\QD)\left(1+l(\QD)\right)^{-1}= r$, 
is obtained exactly where the desirable wholesale price is $w=c/{\cal X}(\QD)$.
\subsection{Both bounds of price of anarchy}
From the above discussion, it turns out that ${\rm PoA}$ in the two distributed cases that the retailer is the leader in push serial supply chain and  the manufacturer is the leader in pull serial supply chain, respectively, namely the inventory is stocked at the leader's site, is similar to the derived results of original newsvendor problem. Herein the others are discussed.

Firstly we explain the case that the manufacturer is the decision maker in push serial supply chain. Utilizing the definition of increasing generalized failure rate,  
$\log{\cal X}(\xi)=\log{\cal X}(0)-\int_0^\xi dy\f{g(y)}{y}$ 
is assessed. Thus with respect to the class of increasing generalized failure rate distribution, $
\max\left({\cal X}(\QC),{\cal X}(\QD)(\QD)^s\xi^{-s}\right)
\le {\cal X}(\xi)
\le
\max\left({\cal X}(\QC),{\cal X}(\QD)(\QD)^k\xi^{-k}\right)
$ is derived where $\QD\le\xi\le\QC$ and $k=g(\QD)$ and $s=g(\QC)$ are represented, respectively. According to the ratio $\a=\QC/\QD$, one can analyze an upper bound and a lower bound of the integration as follows; 
\bea
&&{\cal L}(\a,s)\le\int_{\QD}^{\QC}d\xi{\cal X}(\xi)\nn
&\le& \left\{\begin{array}{ll}\QD{\cal X}(\QD)\f{\a^{1-k}-1}{1-k}&(1-k)^{-\f{1}{k}}\le\a\\
{\cal L}(\a,k)&(1-k)^{-\f{1}{s}}\le\a\le(1-k)^{-\f{1}{k}}
\end{array}\right.\quad\nonumber
\eea
where 
${\cal L}(\a,t):=\QD{\cal X}(\QD)\left[(1-k)\a+\f{t(1-k)^{1-\f{1}{t}}-1}{1-t}\right]
$ 
is employed. Moreover 
$\QD{\cal X}(\QD)\le\int_0^{\QD}d\xi{\cal X}(\xi)
\le \QD{\cal X}(\QD)
\left[1+\f{k}{1-k}\left(1-\left(\f{{\cal X}(\QD)}{{\cal X}(0)}\right)^{\f{1}{k}-1}\right)\right]$ 
is calculated since 
${\cal X}(\QD)\le{\cal X}(\xi)\le\min\left(
{\cal X}(0),{\cal X}(\QD)(\QD)^k\xi^{-k}
\right)$ 
is obtained in $\xi\le\QD$. Therefore an upper bound and a lower bound of price of anarchy in generalized newsvendor problem are evaluated as follows;
\bea
{\rm PoA}&\le&\left\{\begin{array}{ll}
(1-k)^{-\f{1}{k}}-(1-k)^{-1}&(1-k)^{-\f{1}{k}}\le\a\\
\f{\a^{1-k}-(1-k)^2\a}{k(1-k)}-(1-k)^{-1}&{\rm otherwise}
\end{array}\right.\nonumber\ \ \\
{\rm PoA}&\ge&
\f{\dis{\f{s(1-k)^{1-\f{1}{s}}-s}{1-s}-\f{\dis{k\tilde{r}^{\f{1}{k}-1}(1-k)^{1-\f{1}{k}}-k}}{\dis{1-k}}}}{k+\f{\dis{k-k\tilde{r}^{\f{1}{k}-1}(1-k)^{1-\f{1}{k}}}}{\dis{1-k}}}\nonumber
\eea
where we replace the ratio $r$ as the rescaled ratio of the purchasing cost to the selling price, $\tilde{r}:=\f{c}{p{\cal X}(0)}\in[0,1]$. It turns out that the derived bounds in generalized newsvendor problem are as well as the ones in original newsvendor problem. Lastly, both bounds are also simply derived in the case that the retailer is the leader in pull serial supply chain.
\subsection{Example: A toy model}
The previous argument has indicated only if one validates both bounds of the performance ratio in resolving generalized newsvendor problem, we need not to restrict to the literature of supply chain management. As a matter of course, with respect to the ensemble of increasing generalized failure rate without the context of operations management, one also needs to vindicate the improved and developed bounds. Hence we would apply ${\cal M}(Q):=\tanh(Q)$ for simplicity of the discussion because ${\cal X}(Q)=1-\tanh^2(Q)$, $g(Q)=2Q\tanh(Q)$ and $l(Q)={2}{\sinh^2(Q)}$ are derived briefly and analytically.

\begin{figure}[h]
\begin{center}
\includegraphics[width=8cm,height=6cm]{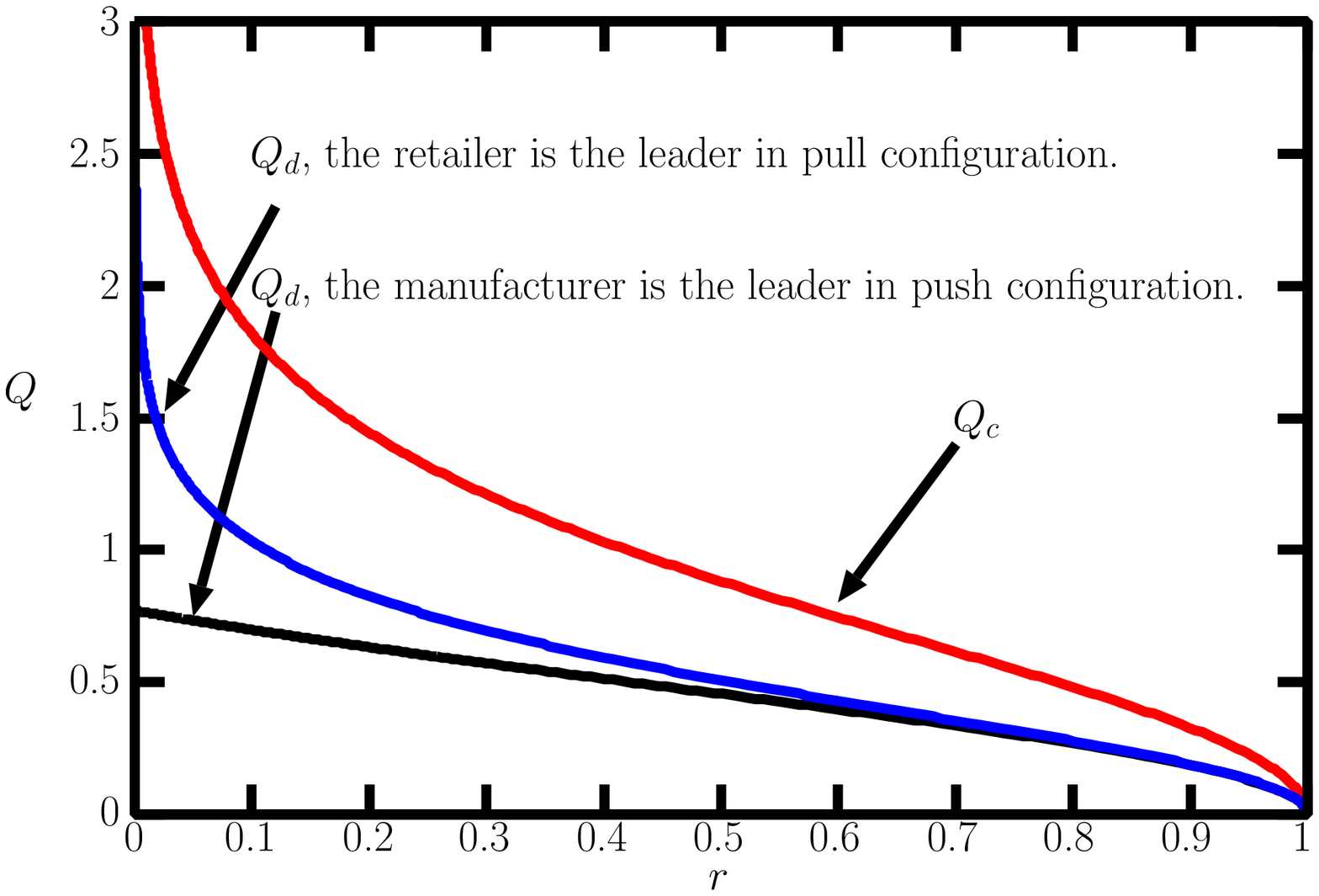}
\caption{\label{tanhpoa1}The ratio $r=c/p$ v.s. $\QC$ and $\QD$.}
\includegraphics[width=8cm,height=6cm]{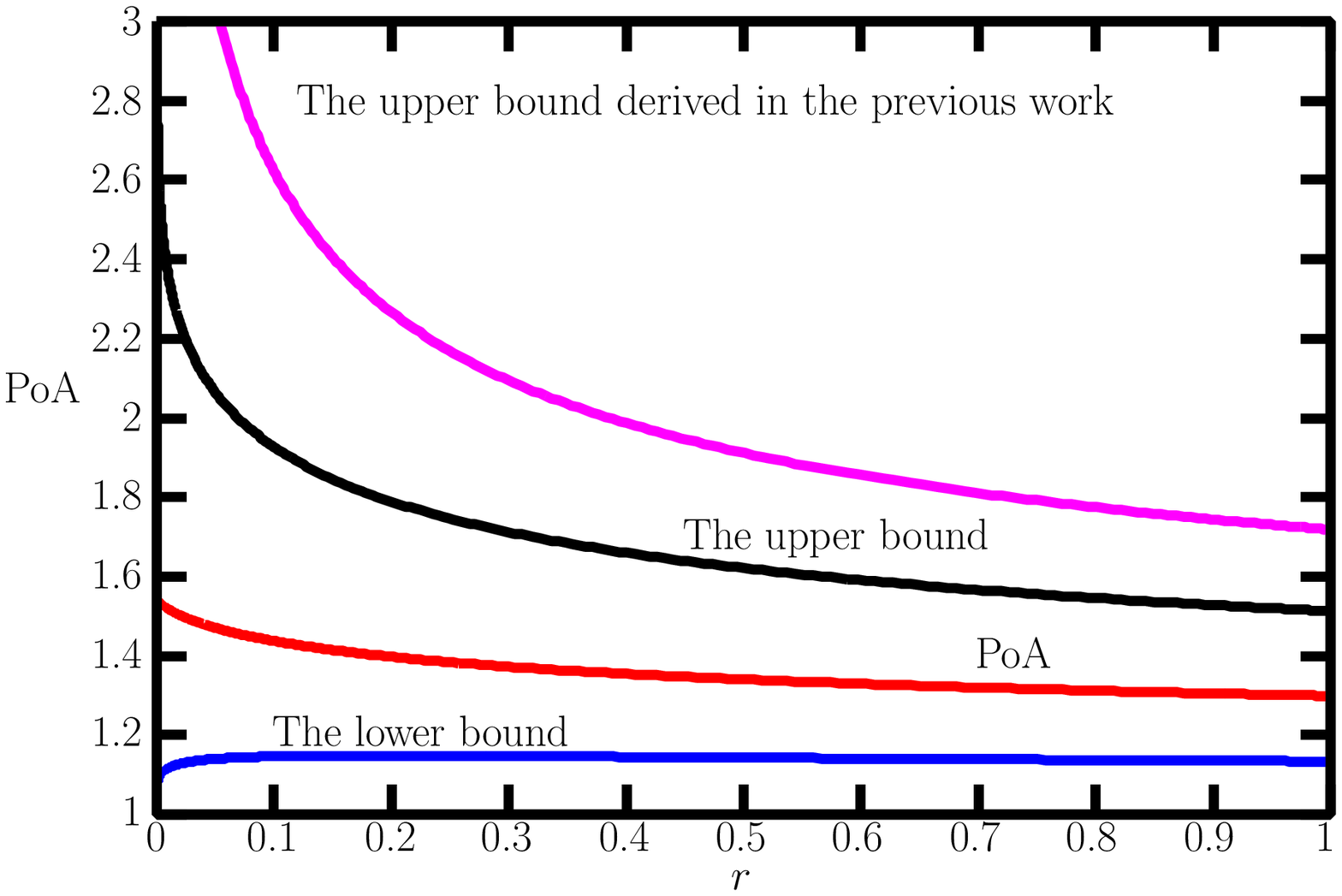}
\caption{\label{tanhpoa2}The ratio $r$ v.s. price of anarchy and the bounds. In push serial supply chain, the manufacturer is the decision maker.}
\includegraphics[width=8cm,height=6cm]{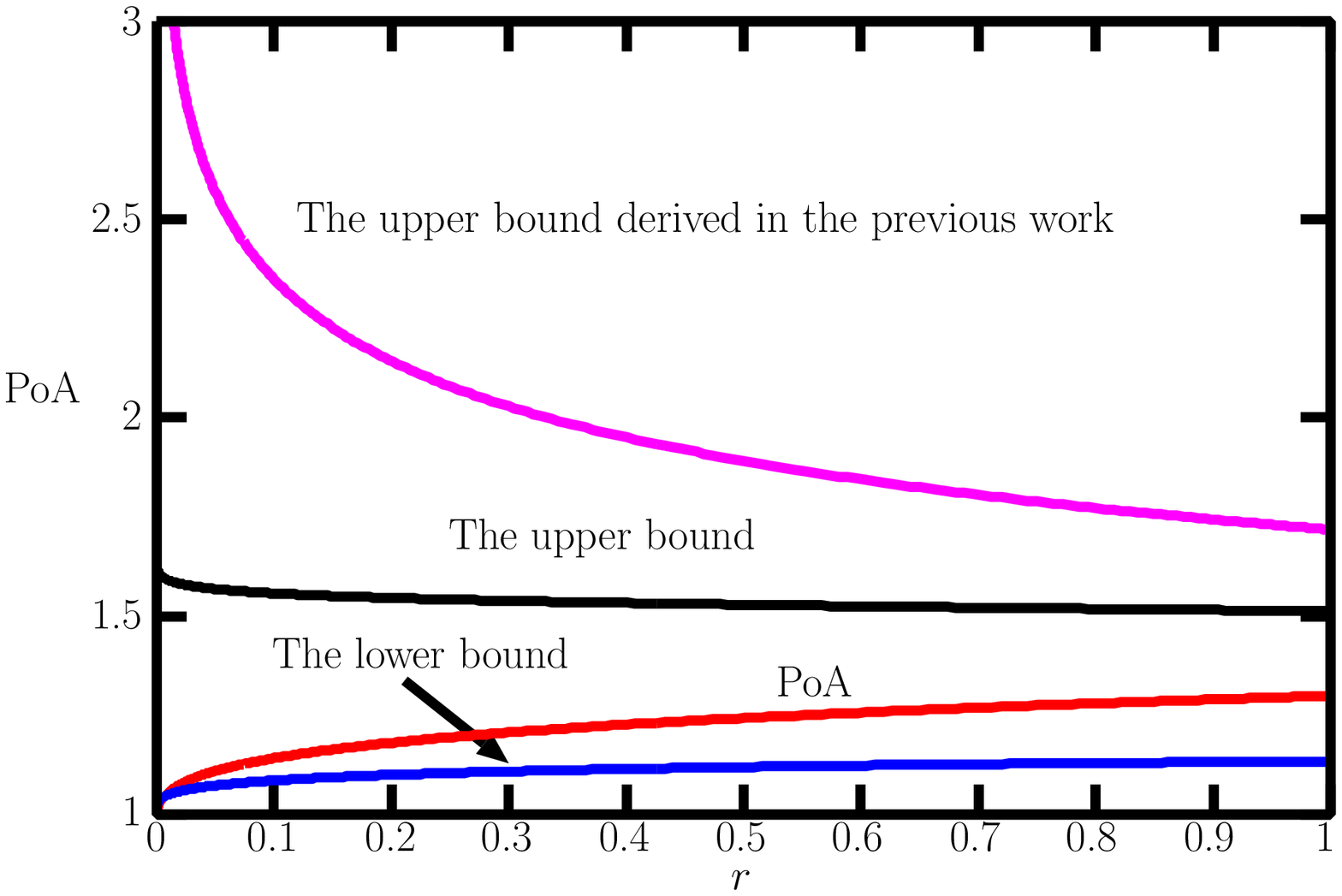}
\caption{\label{tanhpoa3}The ratio $r$ v.s. price of anarchy and the bounds. In pull serial supply chain, the retailer is the decision maker.}
\end{center}
\end{figure}

Firstly, in the case that the manufacturer is the leader in push serial supply chain (as a matter of convenience we address so), it is comparatively easy to execute the algorithm based on \Sref{eq48}, then $\QC$ and $\QD$ are illustrated in \Href{tanhpoa1} and ${\rm PoA}$ and the derived bounds are indicated in \Href{tanhpoa2}. Conclusionally, it turns out that our approach is valid in this case. While, in the case that the retailer is the leader in pull serial supply chain, it is also to perform the iteration based on \Sref{eq52}, then 
$\QC$ and $\QD$ are shown in \Href{tanhpoa1} and ${\rm PoA}$ and the derived bounds are presented in \Href{tanhpoa3}. Similarly it turns out that our procedure is supported validly in this case.
\section{Geometric interpretation\label{sec4}}
In the previous section, we have examined price of anarchy with respect to generalized newsvendor problem. Furthermore here, in order to comprehend the measure in depth \cite{Correa}, let us provide a novel geometric interpretation with respect to price of anarchy via the property of the convexity of $-{\cal M}\left(Q\right)$ \cite{Boyd,Rockafeller}.
\subsection{Geometric interpretation; the integrated supply chain}
\begin{figure}[h]
\begin{center}
\includegraphics[width=8cm,height=6cm]{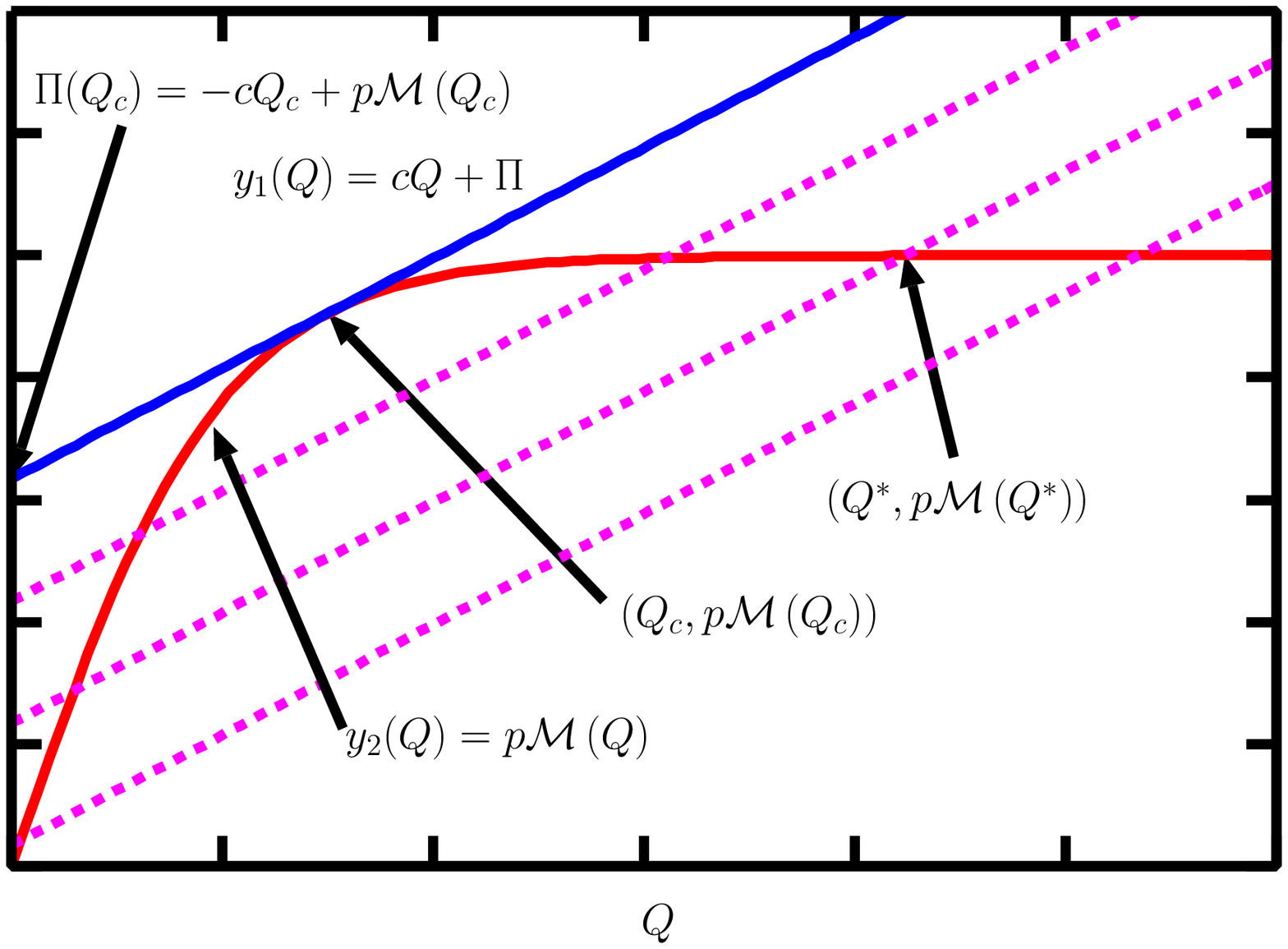} 
\caption{\label{topo2}The maximum of intercept of linear function implies the extremum of the expected total profit function.} 
\end{center}
\end{figure}

First, one can divide two distinguished functions with respect to the expected aggregate profit function, $\Pi=-cQ+p{\cal M}\left(Q\right)$ as follows;
$y_1\left(Q\right)=cQ+\Pi$ and $y_2\left(Q\right)= p{\cal M}\left(Q\right)$ 
where from $y_1(Q)=y_2(Q)$, that is, if there exists intersection point, then $\Pi=-cQ+p{\cal M}\left(Q\right)$ is derived. As shown in \Href{topo2} that $\Pi$ implies the intercept of $y_1(Q)$. Or as another representation, the linear function of $Q$ which has slope $c$ and passes through an intersection point $\left(Q^*,p{\cal M}\left(Q^*\right)\right)$ is represented as follows; $
y_1^*\left(Q\right)=c\left(Q-Q^*\right)+p{\cal M}\left(Q^*\right)$.
Indeed the intercept of $y_1^*(Q)$ describes also $\Pi$. Show \Href{topo2}, in order to maximize the intercept of $y_1(Q)$, both functions should intersect at one point (denoted by $\left(\QC,p{\cal M}\left(\QC\right)\right)$) at least. Since the derivatives of both functions of $Q$, i.e. 
$\pp{y_1\left(Q\right)}{Q}=c$ and $\pp{y_2\left(Q\right)}{Q}
=p{\cal X}\left(Q\right)$, are yielded easily, $\dis{\pp{y_1\left(Q\right)}{Q}=\pp{y_2\left(Q\right)}{Q}}$ at $Q=\QC$, namely
\bea
\label{eq218}c\eq p{\cal M}\left(\QC\right),
\eea
is possible to be sufficiently satisfied. Thus we can resolve the unique optimal solution of the integrated inventory management. 

One point should be worthy to be noticed here. ${\cal X}\left(Q\right)$, the derivative of ${\cal M}\left(Q\right)$ of $Q$, is not always necessary for the continuous function of $Q$ (however by definition, ${\cal M}(Q)$ is satisfied with the continuous function because of the concavity). For example, the given function, 
\bea
p{\cal M}\left(Q\right):=\left\{\begin{array}{ll}
\log\left(1+Q\right)&0\le Q\le \QV\\
v\left(Q-\QV\right)+\log\left(1+\QV\right)&\QV<Q
\end{array}
\right.
\nonumber
\eea
is defined with constant $\QV$ and $v$ where $v<\f{1}{1+\QV}$ is required because ${\cal X}(Q)$ is a nonincreasing function of $Q$. Then with respect to the slope of $y_1(Q)$ in $v\le c\le\f{1}{1+\QV}$, it turns out that the intersection point is $\left(\QV,p{\cal M}\left(\QV\right)\right)$, however, this optimal solution is not satisfied with \Sref{eq218} indeed. That is, we need to comprehend that \Sref{eq218} describes the sufficient condition but not the necessary. Without the loss of generality, so as to prevent also us from misleading in practice, we should confirm the behaviors of both functions $y_1(Q)=cQ+\Pi$ and $y_2(Q)=p{\cal M}(Q)$ being supported by the picture such as \Href{topo2}.

 \begin{figure}[t]
 \begin{center}
\includegraphics[width=8cm,height=6cm]{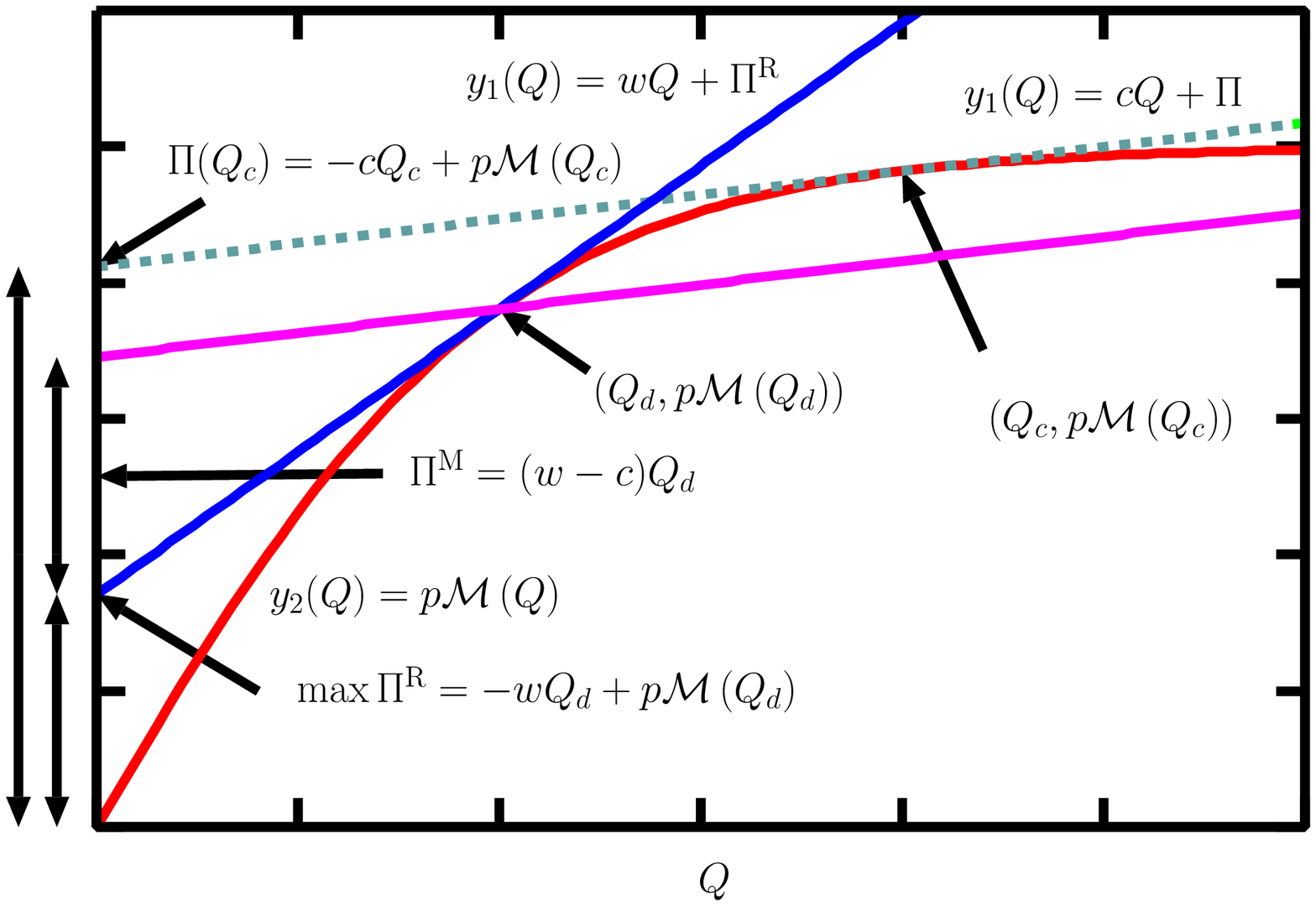} 
\caption{\label{topo1}A geometric interpretation of price of anarchy; the manufacturer is the leader in push serial supply chain. 
\Href{chitopo} illustrates how $\QD$ is determined.}
\includegraphics[width=8cm,height=6cm]{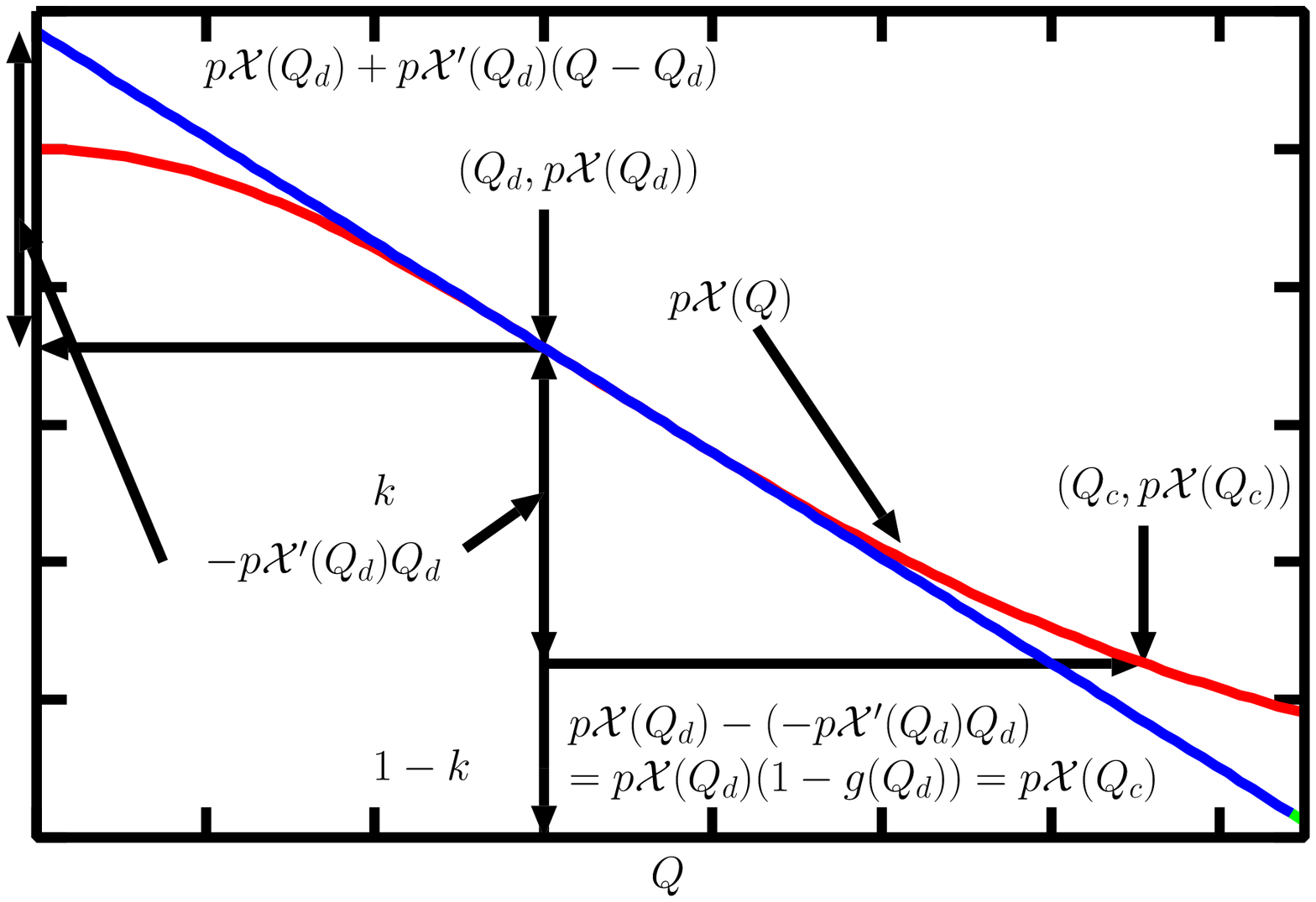} 
\caption{\label{chitopo}$\QD$ is satisfied with $p{\cal X}(\QD)\left(1-g(\QD)\right)=p{\cal X}(\QC)$.}
\end{center}
\end{figure}
\begin{figure}[t]
 \begin{center}
\includegraphics[width=8cm,height=6cm]{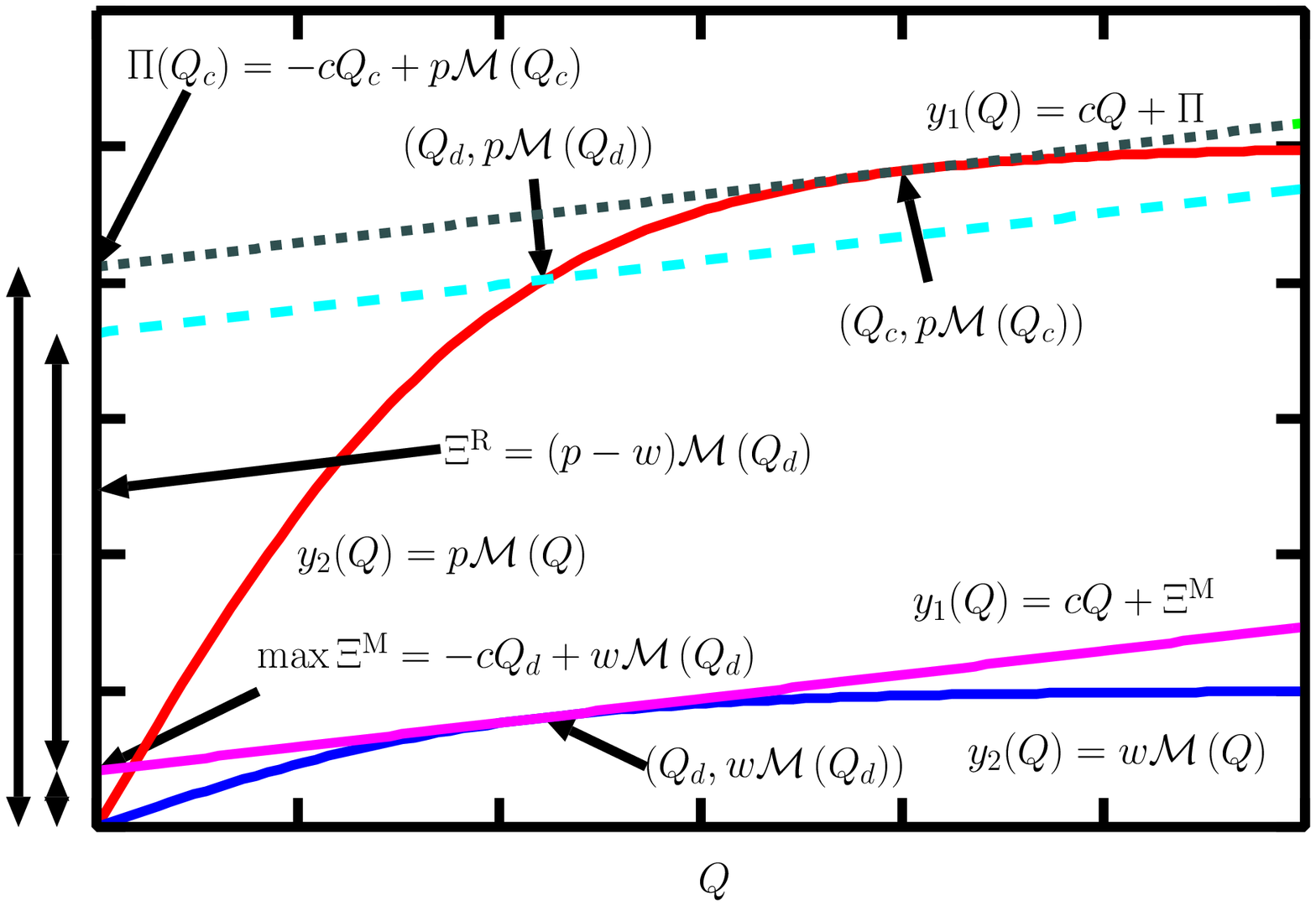} 
\caption{\label{topo4}A geometric interpretation of price of anarchy; the retailer is the leader in pull serial supply chain. \Href{chitopo2} illustrates how $\QD$ is determined.}
\includegraphics[width=8cm,height=6cm]{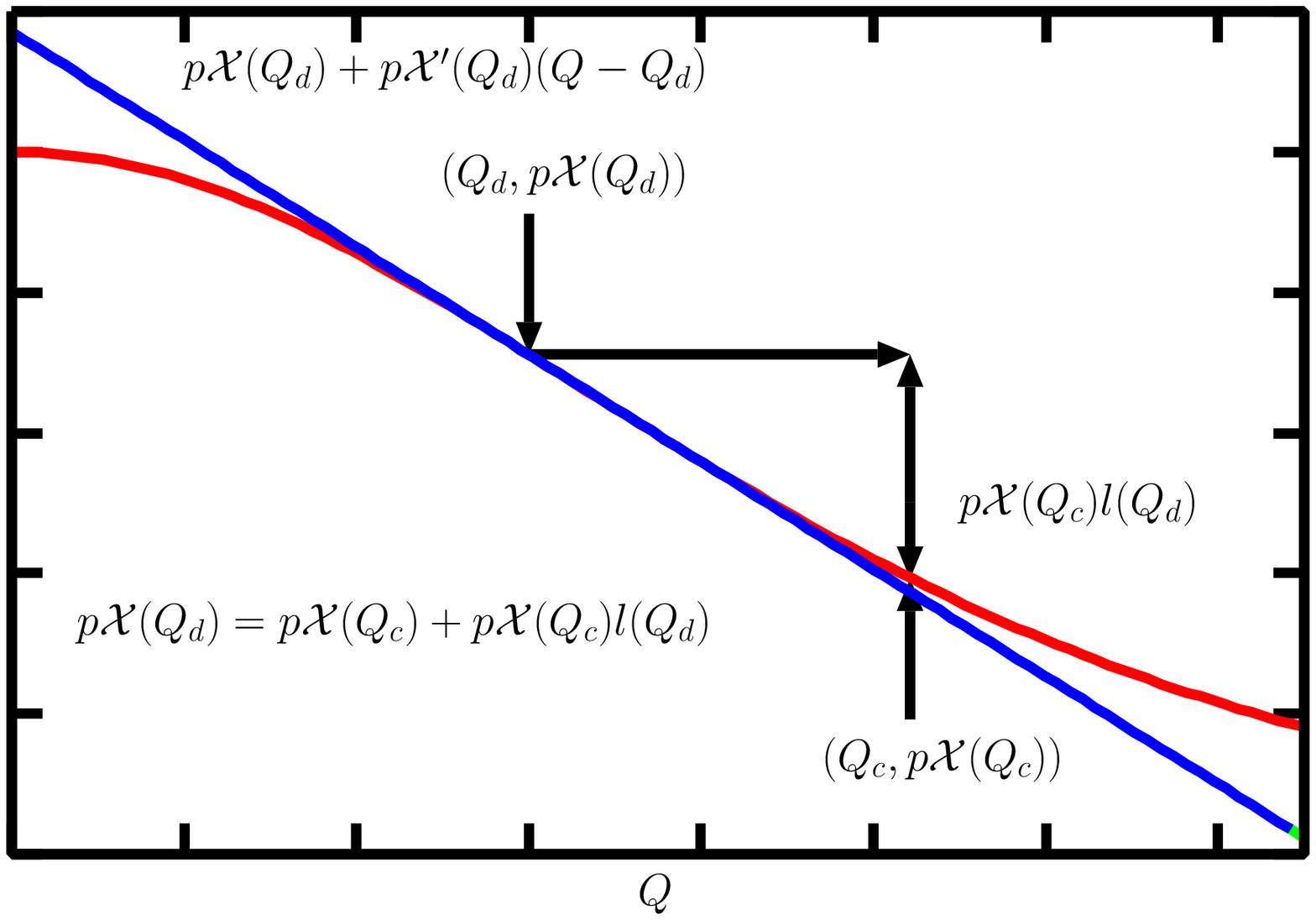} 
\caption{\label{chitopo2}
$\QD$ is satisfied with $p{\cal X}(\QD)\left(1+l(\QD)\right)^{-1}=p{\cal X}(\QC)$.}
\end{center}
\end{figure}
\subsection{Geometric interpretation; the manufacturer is the leader in push serial supply chain}
According to the explanation in the previous subsection, with respect to $\Pi^{\rm R}=-wQ+p{\cal M}(Q)$, in order to solve the optimization problem of the follower, the two functions, $
y_1(Q):=wQ+\Pi^{\rm R}$ and $y_2(Q):=p{\cal M}(Q)$ are denoted. As shown in \Href{topo1} that the supremum of the intercept of $y_1(Q)$ (the optimal solution is $\left(\QD,p{\cal M}\left(\QD\right)\right)$) describes the desirable whole profit and the expected total profit in the centralized case, $\Pi\left(\QC\right)$ is greater than 
$\max\Pi^{\rm R}$ because of $c\le w$. Furthermore since the leader's aggregate profit is represented as $\Pi^{\rm M}=(w-c)\QD$, \Href{topo1} illustrates that $
\Pi\left(\QC\right)\ge (w-c)\QD+\left(-w\QD+p{\cal M}\left(\QD\right)\right)=:\Pi\left(\QD\right)$, where $\Pi(\QD)$ connotes the expected profit of the decentralized case. 
Conclusionally, it turns out that PoA is greater than unit in nature.

\subsection{Geometric interpretation; the retailer is the leader in pull serial supply chain}
In the other case, so as to analyze the optimization problem of the follower, with respect to $\Xi^{\rm M}=-cQ+w{\cal M}(Q)$, the two functions, $
y_1(Q):=cQ+\Xi^{\rm M}$ and $y_2(Q):=w{\cal M}(Q)$, are defined. As shown in \Href{topo4} that the  maximum of the intercept of $y_1(Q)$ (the optimal solution is $\left(\QD,p{\cal M}\left(\QD\right)\right)$) represents the desirable entire profit and the expected aggregate profit in the integrated case $\Pi\left(\QC\right)$ is greater than $\max\Xi^{\rm M}$ because of $w\le p$. Moreover if the leader's whole profit is $\Xi^{\rm R}=(p-w){\cal M}(\QD)$, \Href{topo4} depicts, $\Pi\left(\QC\right)\ge-c\QD+w{\cal M}\left(\QD\right)+(p-w){\cal M}\left(\QD\right)=\Pi\left(\QD\right)$. In conclusion, it turns out that PoA is greater than unit in general.
\vspace{-0.3cm}
\section{\label{sec5}Conclusions}
We discussed price of anarchy, the performance ratio,  which could characterize the loss of efficiency of the distributed supply chain management compared with the integrated supply chain management via newsvendor problem and the generalization instead of bullwhip effect. In particular, the performance ratios in some of remarkable decentralized supply chain managements  are analyzed theoretically and numerically. Furthermore, with respect to the ensemble of increasing generalized failure rate, which one can ensure that the optimization problem of the follower could possess the well-defined solution; (a) the upper bound which has been investigated in the previous work \cite{Perakis-Roels} is improved in this paper utilizing the more accurate evaluation of the integration and (b) the lower bound is derived in the same manner, in the case that the manufacturer could control the wholesale price in push serial supply chain and in the case that the retailer could adjust the wholesale price in pull serial supply chain, namely the two cases that the follower makes to stock the inventory. Moreover the framework handled in section \ref{sec2} has been developed and deepened for generalized newsvendor problem, and we indicate that the loss of efficiency is measured as well as original newsvendor problem. Hence price of anarchy is possible to be one of the most unbeatable feature quantity with respect to the convex optimization involving Stackelberg leadership game \cite{Roughgarden}. While our approach is supported validly in some examples which are satisfied with increasing generalized failure rate. Lastly without the loss of generality, a geometric interpretation of price of anarchy has been provided concretely.

The investigations of geometric interpretation of price of anarchy in multiechelon case, and of the other ensembles which are guaranteed that the optimization problem of the follower can possess the well-defined solution are promising topics for future works.
\section*{Acknowledgment}
One of the authors (TS) appreciates T. Kamishima who works in Advanced Industrial Science and Technology (AIST) for his fruitful advice. 

\appendices
\section{Preliminaries}
\subsection{Global optimal and local optimal\label{appa}}
In this appendix, let us introduce the relationship between global optimal and local optimal. Well, we assume that ${\cal X}$ and ${\cal Y}$ are convex sets and employ as $x\in{\cal X}$ and $y\in{\cal Y}$, respectively. Furthermore two real-valued functions $f(x,y)$ and $g(x,y)$ are bounded from above in the region $(x,y)\in{\cal X}\otimes {\cal Y}$. Then a novel function is denoted as follows; $
F(x,y):=f(x,y)+g(x,y)$ where this function is also satisfied with bounded above and let $(x^*,y^*)$ be an extremal solution of $F(x,y)$ in the given finite region. While $(x^{**},y^{**})$ indicates an extremal solution of $f(x,y)$, that is, one part of $F(x,y)$,  then, $F(x^*,y^*)\ge F(x^{**},y^{**})$ is obtained in general. Thus price of anarchy is greater than or equal to unit by definition. 

\subsection{The concavity of $\int_0^Qd\xi\F(\xi)$\label{appd}\label{appag}}
Because $\F(\xi)$ is a nonincreasing function of $\xi$ firstly, $\int_0^Qd\xi\F(\xi)\le\int_0^{Q_0}d\xi\F(\xi)+\F(Q_0)(Q-Q_0)$ 
is held for any $Q$ and $Q_0$ in general, therefore, for any $Q,Q'$ and $\l\in[0,1]$, the concavity, $\l\int_0^Qd\xi\F(\xi)+(1-\l)\int_0^{Q'}d\xi\F(\xi)
\le\int_0^{\l Q+(1-\l)Q'}\hspace{-1.3cm}\raisebox{-0.4em}{$d\xi\F(\xi)$}$ 
is satisfied when $Q_0=\l Q+(1-\l)Q'$ is rewritten. Additionally, if $Q=\QD$ and $Q_0=\QC$, it is also proved that ${\rm PoA}$ is greater than or equal to unit from 
$\int_0^Qd\xi\F(\xi)\le\int_0^{Q_0}d\xi\F(\xi)+\F(Q_0)(Q-Q_0)$.

\subsection{Young's inequality}
Because $\F(\xi)$ is a nonincreasing function of $\xi$, 
$Q\varphi+\int_{\varphi}^1dy\F^{-1}(y)\ge\int_0^Qd\xi\F(\xi)\ge Q\F(Q)\label{app1}$ 
is held without the loss of generality for any $\varphi\ge0$. The more left inequality is termed as Young's inequality, iff $\F(Q)=\varphi$ gives the equality. Therefore the more right inequality is obtained from $\int_0^Qd\xi\F(\xi)-Q\F(Q)=\int_{\F(Q)}^1dy\F^{-1}(y)\ge0$ at $\F(Q)=\varphi$ without the property of increasing generalized failure rate \cite{Feller,Hardy}.
\subsection{\label{appc}A proof of nondecreasing function $l(Q)$}
If $g(Q)$ is increasing generalized failure rate, then $l(Q)$ is strictly satisfied with 
$\pp{l(Q)}{Q}\ge0$. Because one can prepare a novel function firstly,
$j(Q):
=\f{\dis{1}}{{\cal X}(Q)}\int_0^Qd\xi{\cal X}(\xi)$,
where $\pp{j(Q)}{Q}=1+l(Q)$. Here we allow to rewrite $l(Q)$ as 
$l(Q)=\f{j(Q)g(Q)}{Q}$. 
From the derivative of $\log l(Q)$ with respect to $Q$, 
\bea
\f{1}{l(Q)}\pp{l(Q)}{Q}=\f{1}{g(Q)}\pp{g(Q)}{Q}+\f{\dis{g(Q)-\left(1-\f{Q}{j(Q)}\right)}}{Q},\nonumber
\eea
is calculated where $Q$, $j(Q)$, $g(Q)$ and $l(Q)$ are nonnegative by definition. Hence as proof by contradiction, $g(Q)<1-\f{Q}{j(Q)}$ is assumed. Then the derivative of $1-\f{Q}{j(Q)}$ is derived to be negative exactly. Thus $g(Q)>0>1-\f{Q}{j(Q)}$ is yielded where $\lim_{Q\to0}g(Q)=\lim_{Q\to0}\left(1-\f{Q}{j(Q)}\right)=0$, however this result is inconsistent with the assumption $g(Q)<1-\f{Q}{j(Q)}$, namely $g(Q)\ge1-\f{Q}{j(Q)}$ is held in nature. Therefore it is proved that $l(Q)$ is a nondecreasing function of $Q$.
\subsection{Magnitude relation of price of anarchy}
From Young's inequality and the discussion of the previous appendix, 
$l(Q)\ge g(Q)$ and $l(Q)g(Q)-l(Q)+g(Q)\ge0$ are derived, respectively. Therefore the relationship between the derivative of the inventory level $Q_{d,\rm pull}$ in pull serial supply chain with respect to the rate $r$ and the derivative of the inventory level $Q_{d,\rm push}$ in push serial supply chain with respect to the rate is obtained as $\pp{Q_{d,\rm pull}}{r}\le 
\pp{Q_{d,\rm push}}{r}$ strictly. Thus $Q_{d,\rm pull}=-\int_r^1dr
\pp{Q_{d,\rm pull}}{r}\ge-\int_r^1dr 
\pp{Q_{d,\rm push}}{r}=Q_{d,\rm push}$ is held. Moreover because $\Pi$ is a nondecreasing function of $Q$ in $Q<\QC$, ${\rm PoA}$ in push configuration is greater than or equal to ${\rm PoA}$ in pull configuration in general.
\subsection{Price of anarchy in the fixed order case}
For any $Q\in(0,\QC]$, the sufficient and necessary condition of the equality $1-g(Q)=(1+l(Q))^{-1}$ is ${\cal M}(Q)\propto Q$. In original newsvendor problem, it implies that $\F(\xi)=1$ for $\xi\le Q_0$ and $0$, otherwise, namely the order is fixed at $\xi=Q_0$, then $\QC=\QD=Q_0$ is held in each decentralized supply chain. In conclusion, if the order is constant, the performances of the four cases discussed in this paper are consistent with one another.

\section{$N$ serial supply chain management}
Our approach based on generalized newsvendor problem is simply to be extended $N$ serial supply chain management (the case of $N=2$ is already mentioned). The optimal inventory level of each distributed management is devoted as follows:
\begin{description}
\item[(a)] The manufacturer is the decision maker in push supply chain. 
$\QD$ is derived from the following equation; $\left(1+Q\pp{}{Q}\right)^{N-1}{\cal X}(Q)=r
$ 
where $r=c/p$ and roughly speaking, $\QD$ is probably to be satisfied with the condition ${\cal X}(Q)\left(1-g(Q)\right)^{N-1}\ge r$. Here one point should be noteworthy. The optimal inventory level $\QD$ is not always satisfied with the equality, ${\cal X}(Q)\left(1-g(Q)\right)^{N-1}=r$,  because $\pp{^ng(Q)}{Q^n}=0$ is not always held for each integer in $1<n< N$ Moreover it is hardly desirable that the comparative tight both bounds of price of anarchy in this configuration are derived by the optimal inventory level which is satisfied with the above inequality.
\item[(b)] The retailer is the decision maker in push supply chain.
Since the inventory is stored at the leader's site, $\QD=\QC$ is desirable.
\item[(c)] The manufacturer is the decision maker in pull supply chain.
Nevertheless to say, as the goods is stocked at the upstream site, $\QD=\QC$ is expected.
\item[(d)] The retailer is the decision maker in pull supply chain. $\QD$ is derived as the solution of the following relation,
$\left(1+\f{{\int_0^Qd\xi{\cal X}(\xi)}}{{\cal X}(Q)}\pp{}{Q}\right)^{N-1}\f{1}{{\cal X}(Q)}= \f{1}{r}$ 
where roughly speaking, $\QD$ is possible to be satisfied with the condition ${\cal X}(Q)\left(1+l(Q)\right)^{-(N-1)}\ge r$.
\end{description}

\section{Multiple materials and multiple items}
We could develop our approach in the case of the inventory management of $S$ multiple materials and $I$ multiple items briefly. 
 Let $\vec{c}:=\left\{c_1,c_2,\cdots,c_S\right\}^{\rm  T}\in{\bf R}^S$ and $\vec{Q}:=\left\{Q_1,Q_2,\cdots,Q_S\right\}^{\rm  T}\in{\bf R}^S$ be the purchasing costs and the inventory levels of the materials, respectively. Furthermore $\vec{p}:=\left\{p_1,p_2,\cdots,p_I\right\}^{\rm  T}\in{\bf R}^I$ and $\vec{{\cal M}}
:=\left\{{\cal M}_1
,
{\cal M}_2
,\cdots,{\cal M}_I
\right\}^{\rm  T}\in{\bf R}^I$ represent the selling prices and the order levels of the items, respectively. The entry of the order levels is assumed to be strictly a convex function of $\vec{Q}$. The expected whole profit is defined as follows; $
\Pi=-\vec{c}^{\rm T}\vec{Q}+\vec{p}^{\rm T}\vec{{\cal M}}
$, 
where the notation ${\rm T}$ denotes the matrix transpose. First, the optimal inventory levels of the integrated supply chain $\vec{Q}_c$ is held with the following equation,
\bea
c_s\eq\sum_{\mu=1}^Ip_\mu\left(\pp{{\cal M}_\mu
}{{Q}_s}\right)_{\vec{Q}\to\vec{Q}_c}.\nonumber
\eea
Next, in the case that the manufacturer is the leader in push serial supply chain, the inventory levels of the decentralized configuration $\vec{Q}_d$ is satisfied as follows;
\bea
c_{s}=\sum_{\mu=1}^Ip_\mu\left[
\pp{{\cal M}_\mu
}{Q_s}+
\sum_{t=1}^SQ_t\pp{^2{\cal M}_\mu
}{Q_s\p Q_t}
\right]_{\vec{Q}\to\vec{Q}_d}.\nonumber
\eea
Therefore the wholesale prices $\vec{w}:=\left\{w_1,w_2,\cdots,w_S\right\}\in{\bf R}^S$ is denoted as follows;
$w_s=
\sum_{\mu=1}^Ip_\mu\left(\pp{{\cal M}_\mu
}{Q_s}\right)_{\vec{Q}\to\vec{Q}_d}$.
Lastly, in the case that the retailer is the decision maker in pull serial supply chain, the inventory levels of the distributed configurations $\vec{Q}_d$ is satisfied as follows;
\bea
p_\mu=\sum_{s=1}^Sc_s\left[\pp{Q_s}{{\cal M}_\mu}+\sum_{\nu=1}^I
{\cal M}_\nu\pp{^2Q_s}{{\cal M}_\mu\p{\cal M}_\nu}\right]_{\vec{Q}\to\vec{Q}_d},\nonumber
\eea
where the desirable wholesale prices $\vec{w}$ is held as follows; 
$w_\mu=
\sum_{s=1}^Sc_s
\left(\pp{Q_s}{{\cal M}_\mu
}\right)_{\vec{Q}\to\vec{Q}_d}$.

Three points should be noted here. Firstly, several specific problems which are possible to be handled via our approach have been already investigated \cite{Bernstein,Cachon2001,Lariviere}. Especially, only if the solution with respect to the given optimization problem is definite in several types, utilizing such as Lagrange's multiplier method, we could resolve in the same way. Next, in the case of the supply chain management in multiperiod (i.e. $S=I$), let $c_s$, $p_s$, $Q_s$ and $\xi_s$ be the purchasing cost, the selling price, the inventory level and the demand  at term $s$, respectively. Further the demands $\vec{\xi}:=\left\{\xi_1,\xi_2,\cdots,\xi_S\right\}^{\rm T}$ are distributed with the given multivariate density function $f(\vec{\xi})$ and ${\cal M}_s:=\int_0^\infty d\vec{\xi}f(\vec{\xi})m_s(\vec{\xi},\vec{Q})$ indicates the order at term $s$ where $m_s(\vec{\xi},\vec{Q})$ describes that the order at term $s$ is depended on the market trend strongly. Likewise, it turns out that one can deal with multiperiod case based on our approach (c.f. bullwhip effect). Lastly, in the case of $N$ serial supply chain, the optimal inventory levels $\vec{Q}_d$ in the two cases are satisfied with the following equations;
\bea
c_s
\eq\sum_{\mu=1}^Ip_\mu\left[\mathop{\rm Tr}_{\vec{t}}\prod_{i=1}^{N-1}\left(\d_{t_is}+Q_{t_i}\pp{}{Q_{t_i}}\right)\pp{{\cal M}_\mu}{Q_s}\right]_{\vec{Q}\to\vec{Q}_d},\nn
p_\mu
\eq\sum_{s=1}^Sc_s\left[
\mathop{\rm Tr}_{\vec{\nu}}\prod_{i=1}^{N-1}\left(\d_{\nu_i\mu}+{\cal M}_{\nu_i}\pp{}{{\cal M}_{\nu_i}}\right)\pp{Q_s}{{\cal M}_\mu}
\right]_{\vec{Q}\to\vec{Q}_d},\nonumber
\eea
respectively, where $\vec{t}:=\left\{t_1,t_2,\cdots,t_{N-1}\right\}$ and $\vec{\nu}:=\left\{\nu_1,\nu_2,\cdots,\nu_{N-1}\right\}$ are used, $\d_{ab}$ indicates Kronecker's delta which is the entry of the unit matrix and the notation $\mathop{\rm Tr}_{\vec{t}}$ and $\mathop{\rm Tr}_{\vec{\nu}}$ denote the summation over all possible states of $\vec{t}$ and of $\vec{\nu}$, respectively.

\ifCLASSOPTIONcaptionsoff
  \newpage
\fi

\end{document}